\documentclass[useAMS,usenatbib]{mn2e}
\usepackage{graphicx,widetext}
\usepackage{amssymb, amsmath}
\usepackage{times}
\usepackage{color}
\usepackage{lscape}
\usepackage[section]{placeins}
\usepackage{epstopdf}
\usepackage{epsf}
\usepackage{ulem}
\def\prd{Phys. Rev. D}
\def\apj{ApJ}
\def\aap{A \& A}
\def\mnras{MNRAS}
\def\jcap{JCAP}

\newcommand{\nbar}{\bar{n}}

\newcommand{\DA}{D\!_A(z)}
\newcommand{\hz}{H(z)}
\newcommand{\DAA}{D_A\,}
\newcommand{\hzz}{H\,}

\newcommand{\sDA}{\sigma_\DA}
\newcommand{\sHz}{\sigma_\hz}

\newcommand{\DV}{D_V}
\newcommand{\sDV}{\sigma_{D_V}}
\newcommand{\hMpc}{h^{-1}{\rm\;Mpc}}

\newcommand{\trihGpc}{h^{-3}{\rm\;Gpc^3}}
\newcommand{\trihMpc}{h^{-3}{\rm\;Mpc^{3}}}
\newcommand{\itrihMpc}{h^{3}{\rm\;Mpc^{-3}}}
\newcommand{\ihMpc}{h{\rm\;Mpc^{-1}}}

\newcommand{\alpe}{\alpha_\perp}
\newcommand{\alpa}{\alpha_\parallel}

\newcommand{\Sigxy}{\Sigma_{\rm xy}}
\newcommand{\Sigz}{\Sigma_{z}}

\newcommand{\Sigsm}{\Sigma_{\rm sm}}
\newcommand{\Signl}{\Sigma_{\rm nl}}

\newcommand{\Plin}{P_{\rm lin}}

\newcommand{\hdelta}{\hat{\delta}}

\newcommand{\kperp}{k_\perp}
\newcommand{\kpar}{k_\parallel}

\newcommand{\llan}{\left\langle}
\newcommand{\rran}{\right\rangle}
\newcommand{\kvec}{\vec{k}}

\newcommand{\nPt}{nP_{0.2}}
\newcommand{\Pt}{P_{0.2}}

\newcommand{\Vsur}{V_{\rm survey}}

\newcommand{\ie}{i.e.}

\newcommand{\bx}{{\bf x}}

\newcommand{\bz}{{\bf z}}
\newcommand{\bq}{{\bf q}}
\newcommand{\bk}{{\bf k}}

\newcommand{\enia}[1]{e^{-i #1 }}
\newcommand{\eia}[1]{e^{i #1 }}
\newcommand{\enikq}{e^{-i\bk\cdot\bq}}
\newcommand{\enikx}{e^{-i\bk\cdot\bx}}

\newcommand{\dx}{\delta(\bx)}

\newcommand{\dks}{\tilde{\delta_s}(\bk)}

\newcommand{\dkd}{\tilde{\delta_d}(\bk)}

\newcommand{\bPsi}{{\bf \Psi}}
\newcommand{\bs}{\bf s}
\newcommand{\bPsin}[1]{{\bf \Psi^{#1}}}

\newcommand{\PL}{P_{\rm L}}
\newcommand{\Pnl}{P_{\rm nl}}
\newcommand{\Pmc}{P_{\rm MC}}
\newcommand{\Pmct}{P_{\rm MC,0.2}}

\newcommand{\del}{\delta}
\newcommand{\Prec}{P_{\rm rec}}
\newcommand{\Pdd}{P_{\rm dd}}
\newcommand{\Psd}{P_{\rm sd}}
\newcommand{\Pss}{P_{\rm ss}}
\newcommand{\Sigdd}{\Sigma_{\rm dd}}
\newcommand{\Sigsd}{\Sigma_{\rm sd}}
\newcommand{\Sigss}{\Sigma_{\rm ss}}

\newcommand{\lamo}{\lambda_d}
\newcommand{\lamt}{\lambda_s}
\newcommand{\bea}{\begin{eqnarray}}
\newcommand{\eea}{\end{eqnarray}}

\newcommand{\nnm}{\nonumber}

\newcommand{\indlo}{\int\frac{\lambda_1}{2\pi}}
\newcommand{\indlt}{\int\frac{\lambda_2}{2\pi}}
\newcommand{\tF}{\tilde{F}}
\newcommand{\Pobs}{P_{\rm obs}}
\newcommand{\Pbias}{P_{\rm bias}}

\newcommand{\dko}{\tilde{\delta}_{\rm obs}(\bk)}

\newcommand{\pshot}{p_{\rm shot}}

\newcommand{\bso}{\bs_1}
\newcommand{\bst}{\bs_2}

\newcommand{\Cg}{C_G}
\newcommand{\Cmg}{C_{\rm MG}}

\newcommand{\Sigddxy}{\Sigma_{\rm dd,xy}}
\newcommand{\Sigddz}{\Sigma_{\rm dd,z}}

\newcommand{\Sigssxy}{\Sigma_{\rm ss,xy}}
\newcommand{\Sigssz}{\Sigma_{\rm ss,z}}

\newcommand{\scr}{}

\begin{document}

\setlength{\parskip}{0pt}
\title{Modeling the reconstructed BAO in Fourier space}
\author[Seo et al.]
{
Hee-Jong Seo $^{1}$, Florian Beutler  $^{2}$, Ashley J. Ross  $^{3,4}$, Shun Saito  $^{5}$ \\
$^{1}$ Department of Physics and Astronomy, Ohio University, Clippinger Labs, Athens, OH 45701\\
$^{2}$ Lawrence Berkeley National Lab, 1 Cyclotron Rd, Berkeley CA 94720, USA \\
$^{3}$ Center for Cosmology and Astroparticle Physics, Department of Physics, The Ohio State University, OH 43210, USA\\
$^{4}$ Institute of Cosmology \& Gravitation, Dennis Sciama Building, University of Portsmouth, Portsmouth, PO1 3FX, UK\\
$^{5}$ Kavli Institute for the Physics and Mathematics of the Universe (WPI), \\
The University of Tokyo Institutes for Advanced Study, The University of Tokyo, Chiba 277-8582, Japan
}

\date{\today} 
\maketitle
\label{firstpage}

\begin{abstract}
The density field reconstruction technique, which was developed to partially reverse the nonlinear degradation of the Baryon Acoustic Oscillation (BAO) feature in the galaxy redshift surveys, has been successful in substantially improving the cosmology constraints from recent galaxy surveys such as Baryon Oscillation Spectroscopic Survey (BOSS). We estimate the efficiency of the reconstruction method as a function of various reconstruction details. To directly quantify the BAO information in nonlinear density fields before and after reconstruction, we calculate the cross-correlations (i.e., propagators) of the pre(post)-reconstructed density field with the initial linear field using a mock galaxy sample that is designed to mimic the clustering of the BOSS CMASS galaxies. The results directly provide the BAO damping as a function of wavenumber that can be implemented into the Fisher matrix analysis. We focus on investigating the dependence of the propagator on a choice of smoothing filters and on two major different conventions of the redshift-space density field reconstruction that have been used in literature. 
By estimating the BAO signal-to-noise for each case, we predict constraints on the angular diameter distance and Hubble parameter using the Fisher matrix analysis. We thus determine an optimal Gaussian smoothing filter scale for the signal-to-noise level of the BOSS CMASS.
We also present appropriate BAO fitting models 
for different reconstruction methods based on the first and second order Lagrangian perturbation theory in Fourier space. Using the mock data, we show that the modified BAO fitting model can substantially improve the accuracy of the BAO position in the best fits as well as the goodness of the fits.
\end{abstract}

\begin{keywords}
\end{keywords}

\section{Introduction}
\label{sec:intro}
Baryons acoustic oscillations (hereafter BAO) refer to primordial sound waves that propagated in the hot plasma of photons and baryons due to the interaction of gravity and photon pressure in very early Universe. As the Universe expands and cools down, the photons and baryons decoupled near the epoch of recombination and the propagating sound waves subsequently stalled, leaving a distinct statistical imprint in the distribution of matter on the scale that corresponds to the distance the sound waves have traveled before this epoch (i.e., sound horizon scale). 

The BAO feature in the matter distribution reveals itself in the distribution of galaxies. Its observation has proven to be a powerful and robust dark energy probe (see, e.g., \citealt{DR11}). The comparison between the observed BAO scale from the galaxy surveys and the true physical sound horizon scale that can be independently estimated from the Cosmic Microwave Background data enables mapping between the observational coordinates and the physical coordinates; this mapping sensitively depends on the expansion history and therefore dark energy properties~\citep[e.g.,][]{Eisen98}. The precision to which the BAO feature can be measured, and consequently, the dark energy constraints obtained from BAO measurements depend on the strength of the feature and how well we can separate the BAO feature from the broadband power. Due to structure formation as well as the observational effect of redshift-space distortions, the BAO feature is expected to become gradually degraded/damped with structure evolution such that the feature is much weaker at low redshift where the expansion of the Universe is mostly driven by dark energy \citep[see, e.g.,][]{Jain94,Meiksin99,Springel05,Angulo05,SE05,White05,Jeong06,Crocce06b,ESW07,Huff07,Smith07,Seo08, Angulo08,Crocce08,Mat08a,Mat08b,Taka08,Taruya09,Tassev12a}.

\citet{ESSS} has shown that a simple density field reconstruction scheme based on the linear continuity equation can reverse a significant portion of the nonlinear degradation and reconstruct the BAO feature at low redshift. The method has been extensively tested and has now become a standard tool for analyzing recent galaxy surveys \scr{\citep[e.g.,][]{Seo08,PadLPT,Seo10, Noh2009, Mehta11,Pad12,2012MNRAS.427.3435A,DR11,White15,Cohn16}} (also see \citet{Tassev12b} for optimizations). While the basic procedure of BAO reconstruction is well established, the details that go into reconstruction differ depending on literature. For example, the reconstruction efficiency depends on the choice of the scale that smooths the observed (and therefore noisy) nonlinear density field on small scales in order to make the continuity equation valid \scr{\citep[e.g.,][]{PadLPT,Seo10, Mehta11,Vargas14,Burden14,Vargas15,Achitouv15}}. 

In particular, the original BAO reconstruction technique in~\citet{ESSS} attempts to preserve the redshift-space distortion (RSD) signal  \citep[i.e., observational clustering distortions along the line of sight due to the peculiar velocity field of galaxies, e.g.,][]{Kaiser87} in the reconstructed density field in order to take advantage of the boosted signal relative to the shot noise, while the convention in~\citet{Pad12} removes the RSD signal during reconstruction, attempting to recover the isotropic clustering on large scales. As we will present in this paper, we expect that the latter convention should more effectively sharpen the BAO feature along the line of sight, reversing much of the damping due to nonlinear RSD. The final gain in terms of BAO signal to noise also depends on the effect of the reconstruction process on the component of the power spectrum that does not include the BAO information, which would contribute to noise; this effect has \scr{not} been rigorously investigated.  \citet{Burden14} has tested these two RSD conventions using dispersions of the Baryon Oscillation Spectroscopic Survey (BOSS) Data Release 11 mock catalogs (600 mocks) and found no difference between the two methods, when focusing on the spherically averaged BAO signal. Recently, \citet{Vargas15} has also investigated such BAO reconstruction details in correlation function measurements.  Our analysis is different from \citet{Burden14}  in that we inspect the angle-dependent behavior and our approach should in general suffer less sample variance since the initial sample variance effect largely cancels out when calculating the propagator. Our result will thus quantitatively compare the underlying information content components between the two methods, which expand and improve upon the results observed in \citet{Burden14}. 

In this paper, we estimate the efficiency of the reconstruction method as a function of various reconstruction details using a hybrid of simulations and Fisher matrix analysis. To directly quantify the BAO information in nonlinear density fields before and after reconstruction, we calculate the cross-correlations (propagators) of the pre(post)-reconstructed density field with the initial linear field. We do this using a mock galaxy sample that is designed to mimic the clustering of the BOSS CMASS galaxies \citep[stands for `constant stellar mass galaxies',][]{White11,Reid15,Leauthaud15,Saito15}. Since both the initial density field and the final density field are subject to the same initial random fluctuations, this method suffers little sample variance and does not require a large ensemble of simulations. The results directly provide the BAO damping as a function of wavenumber that can be implemented into the Fisher matrix analysis. We focus on investigating the dependence of the propagator on a choice of smoothing filters and on two different conventions of the redshift-space density field reconstruction mentioned above. We also estimate the noise component without the BAO signal, which we call `mode-coupling term', for post-reconstructed fields. Together we estimate the BAO signal to noise for each case and predict constraints on the angular diameter distance and Hubble parameter using the Fisher matrix analysis. We provide an optimal smoothing filter scale for the signal to noise level of the BOSS CMASS. We compare the final Fisher matrix predictions with the current BOSS CMASS constraints. 

Understanding the differences between the reconstruction conventions is important not only for maximizing the gain given the survey data, but also for correctly modeling the reconstructed BAO feature to be used during the data analysis. For example, the BAO feature from low-redshift galaxy surveys has been traditionally modeled as a Gaussian damping of the BAO feature in the linear power spectrum. The model has been shown to be a good approximation for pre-reconstructed density field and post-reconstructed density field \citep{Seo10} for the reconstruction convention used in ~\citet{ESSS} (i.e., retaining RSD). However, as we will show in this paper, the Gaussian damping model is a poor approximation for the density field reconstructed using the convention in ~\citet{Pad12} (i.e., removing RSD) especially if we are interested in the clustering along the line of sight in redshift space. The first order Lagrangian perturbation theory (LPT) model for the reconstructed BAO feature in real space has been derived in \citet{PadLPT} and subsequently extended in \citet{Noh2009} to include the second order terms as well as the effects of galaxy bias. Recently, \citet{White15} has derived the full first order LPT model for the reconstructed BAO in redshift-space, for the correlation function and for both RSD conventions. We follow \citet{Mat08b} and conduct a corresponding first and second order LPT derivation for the BAO feature of the reconstructed power spectrum and provide an effective, simple model for both conventions. For the reconstruction convention based on  ~\citet{Pad12}, our model requires a modification to the original Gaussian damping model.

We show that using the modified model significantly improves the $\chi^2$ when compared to mock datasets. Without this modified model we obtain large values of $\chi^2$, which could impact our judgement of the goodness of fit, and also obtain systematically biased BAO measurements. We will also show that using the modified model can potentially improve the precisions of the measurement.
  
This paper is structured as follows. In \S~\ref{sec:recon}, we outline the density field reconstruction method and the details of the different conventions we investigate in this paper. In \S~\ref{sec:mock}, we summarize the mock simulation we use. In \S~\ref{sec:modelBAO}, we introduce two models of the BAO damping in the reconstructed density field as a single element (the propagator)  and discuss estimators of the noise element (which we call the mode-coupling term). In \S~\ref{sec:results-signal} and \S~\ref{sec:results-noise}, we present the results of the mock analysis in terms of the BAO signal and noise, that is, the propagators and mode-coupling components as a function of the smoothing scale and the reconstruction conventions.  In \S~\ref{sec:discuss}, based on the simulated propagators and mode-coupling contributions, we conduct the Fisher matrix analysis to forecast BAO measurement errors in various cases. We also conduct the BAO fitting on the mock simulation and compare with the Fisher forecasts. In \S~\ref{sec:con}, we conclude. In the Appendix, we provide the first and second order LPT derivations of the reconstructed BAO in power spectra in redshift space, which motivate our modified BAO fitting model.

\section{Density field reconstruction}\label{sec:recon}
\subsection{Basics}
The outline of the density field reconstruction technique (hereafter, BAO reconstruction) is presented originally in \citet{ESSS} and a more detailed description can be found in \citet{Mehta11}. We briefly summarize it here.
The Eulerian particle position $\bx$ can be mapped to a Lagrangian particle position $\bq$ by the displacement $\bPsi$:
\begin{eqnarray}
\bx&=&\bq+\bPsi(\bx).
\end{eqnarray}
The density field, $\delta$, in Eulerian coordinates is transformed to the uniform density field in Lagrangian coordinates through the Jacobian, and the transformation in the linear order gives the familiar continuity equation: 
\begin{eqnarray}
\delta(\bx) &=& - \nabla_\bx \cdot \bPsi(\bx),
\end{eqnarray}
which in Fourier space can be expressed as
\begin{eqnarray}
\tilde{\bPsi}(\bk) &=& - \frac{i\bk}{k^2} \tilde{\delta}(\bk),
\end{eqnarray}
where $\bk$ is the wave vector. 

The idea of BAO reconstruction is to estimate the displacement field due to structure growth and redshift-space distortions based on the observed galaxy density field and displace the observed galaxies back to their original positions to restore the linear information. 

In justification of estimating the displacement field using the linear continuity equation, we damp the nonlinearities on small scales in the observed density field by applying the smoothing kernel $S(k)$ to the density field. Then, the estimated displacement field $\tilde{\bf s}(\bk)$ is:
\begin{eqnarray}\label{eq:qdiv}
\tilde{\bf s}(\bk) &\approx& - \frac{i\bk}{k^2} \tilde{\delta}_{\rm nl} S(k),
\end{eqnarray}
where $S(k)$ is traditionally given by
\begin{eqnarray}
S(k) =  \exp{\left[ -0.25k^2\Sigsm^2\right]}.\label{eq:sk}
\end{eqnarray}
The impact of the smoothing scale $\Sigsm$ will be discussed in detail later. Note that we follow the smoothing scale convention from \citet{PadLPT}; the smoothing scale convention $\Sigsm'$ in \citet{2012MNRAS.427.3435A} as well as \citet{Burden14} and \citet{Vargas15} is different from ours in that $\Sigsm=\sqrt{2}\Sigsm'$.  

\subsection{Isotropic and Anisotropic BAO reconstructions}\label{subsec:anirec}
In this subsection we compare the two main BAO reconstruction conventions that have been used in the literature. The original BAO reconstruction technique in~\citet{ESSS,Seo08,Seo10,Mehta11} attempts to keep the redshift-space distortion (RSD) signal in the reconstructed density field, while the convention of~\citet{Pad12} and~\citet{DR11} attempts to remove the RSD signal during the reconstruction. 
We term the former BAO convention that restores RSD ``anisotropic BAO reconstruction" or ``Rec-Ani'' and the latter convention that removes RSD ``isotropic BAO reconstruction" or ``Rec-Iso".

Parameters that define the impact of RSD on the galaxy field include
\begin{equation}
f \equiv \frac{{\rm dlog}(D)}{{\rm dlog}(a)},
\end{equation}
where $D$ is the linear growth factor and $a$ is the scale factor ($a\equiv1/(1+z)$). On linear scales, RSD enhance the galaxy density field by \citep{Kaiser87}
\begin{equation}
\tilde{\delta}^s_g= (b+f\mu^2)\tilde{\delta}^{r}_m=b (1+\beta\mu^2)\tilde{\delta}^r_m,
\end{equation}
where $b$ is the linear galaxy bias, $\tilde{\delta}^r_m$ is the real space dark matter density field as a function of wave vector, $\tilde{\delta}^s_g$ is the redshift space galaxy density field and $\mu$ is the cosine of the angle between the line of sight and the wave vector. It is useful to define the quantity $\beta = f/b$.

\begin{itemize}
\item {\bf Isotropic BAO reconstruction -- ``Rec-Iso'' }~\citep[e.g.,][]{Pad12,DR11} \\
The real-space displacement field $\tilde{\bf s}^r(\bk)$ is estimated from the observed redshift-space density field $\tilde{\delta}^s_{\rm nl}(\bk)$ by;
\begin{eqnarray}
&&\tilde{\bf s}^r(\bk) = - \frac{i\bk}{k^2} \frac{\tilde{\delta}^s_{\rm nl}(\bk)}{b(1+\beta\mu^2)}S(k).\nonumber\\
\label{eq:recisoq}
\end{eqnarray}

The displacement field in configuration space is then derived by Fourier-transforming $\tilde{\bf s}^r(\bk)$;
\begin{eqnarray}
&&\tilde{\bf s}^r(\bk) \xrightarrow{\rm Fourier Transform} {\bf s}^r(\bx).\\
\end{eqnarray}
The two density fields $\delta_d(\bx)$ and $\delta_s(\bx)$ are then derived, respectively by;
\begin{eqnarray}
&&\delta_d : \mbox{displacing the galaxies by } {\bf s}^s={\bf s}^r + f ({\bf s}^r\cdot\hat{\bz})\hat{\bz}\nonumber\\
&&\delta_s: \mbox{displacing the random particles by } {\bf s}^s={\bf s}^r , \label{eq:reciso}
\label{eq:recisoqt}
\end{eqnarray}
where $\hat{\bz}$ is the unit vector pointing along the line of sight and $f$ is the growth rate. 

\item {\bf Anisotropic BAO reconstruction -- ``Rec-Ani" }~\citep[e.g.,][]{ESSS,Seo08,Seo10,Mehta11}\\
The procedure is similar except for;
\begin{eqnarray}
&&\tilde{\bf s}^{r}(\bk) = - \frac{i\bk}{k^2} \frac{\tilde{\delta}^s_{\rm nl}(\bk)}{b}S(k)\nonumber\\
&&\tilde{\bf s}^r(\bk) \xrightarrow{\rm Fourier Transform} {\bf s}^r(\bx)
\end{eqnarray}
\begin{eqnarray}
&&\delta_d : \mbox{displacing the galaxies by } 
{\bf s}^s=  {\bf s}^r+ \frac{f-\beta}{1+\beta}({\bf s}^r\cdot\hat{\bz})\hat{\bz}\nonumber\\
&&\delta_s : \mbox{displacing the random particles by }
{\bf s}^s=  {\bf s}^r+ \frac{f-\beta}{1+\beta}({\bf s}^r\cdot\hat{\bz})\hat{\bz}\nonumber\\ \label{eq:recani}
\end{eqnarray}
\end{itemize}
The final reconstructed field is derived by 
\begin{eqnarray}
\delta_{\rm rec}(\bx) = \delta_d(\bx)-\delta_s(\bx).\label{eq:drec}
\end{eqnarray}
As discussed in \citet{Pad12}, $\delta_d$ has a restored small-scale information and $\delta_s$ has a restored large-scale information. 

To the lowest order, the displacement field that is responsible for the observed redshift-space density field  can be thought of $\sim -(1+f)/(1+\beta)\frac{i\bk}{k^2} \frac{\hat{\delta}^s_{\rm nl}}{b}S(k)$ along the line of sight and $\sim -1/(1+\beta)\frac{i\bk}{k^2} \frac{\hat{\delta}^s_{\rm nl}}{b}S(k)$ across the line of sight.
Note that both conventions described above displace the real galaxy particles approximately by this estimated redshift-space displacement field, thereby attempting to correct for RSD on small scales, while each convention makes a different choice on displacing random (reference) particles. 
The anisotropic BAO reconstruction displaces random particles with the estimated redshift-space displacement field thereby boosting the large-scale clustering in $\delta_s$ along the line of sight, while the isotropic BAO reconstruction displaces the random particles without the extra factor of $(1+f)$, attempting to correct for the anisotropic redshift distortions on large scales. 
While the former has the boosted amplitude along the line of sight, the latter has effectively a sharper BAO peak along the line of sight.

\section{Mock data}\label{sec:mock}
We utilize the mock galaxy data (hereafter, `runA' mocks) used in \citet{White11} that mimic the observed galaxy clustering of the BOSS-CMASS data.  We use this mock simulation, first to estimate the cross-correlations between the initial and the final density fields (i.e., propagators) for various cases and later to simulate the BAO fitting using different fitting models. The BOSS CMASS data correspond to a signal to noise ratio of $\nPt \sim 1.8$ where $n$ is the average number density, $1/n$ is the corresponding shot noise estimate, and $P_{0.2}$ is the amplitude of power at $k=0.2\ihMpc$ without redshift-space distortions. As a caveat, the assumption of $\nPt \sim 1.8$ breaks down at the very high redshift end of the BOSS CMASS data, as  shown in Table 2 of \citet{DESI}.
The mock contains 1,203,407 mock galaxies within a cosmic volume of $1.5^3\trihGpc$, i.e., $\nbar=3.566\times 10^{-4} \itrihMpc$,
which gives a shot noise level of $2804\trihMpc$ and $\nPt\sim 2$, i.e., close to the signal to noise of the BOSS CMASS data \footnote{Note that \citet{Burden14} assumes a somewhat lower signal to noise for BOSS CMASS. They assume the weighted effective number density $\bar{n}=2.196 \times 10^{-4} \itrihMpc$, i.e., a shot noise level of $4553 \trihMpc$.}.
The redshift of the mock galaxies is 0.55 ( for the BOSS CMASS data, $z_{\rm eff}=0.57$). 
We assume $b=2$ and $f=0.74$ for the analysis in this paper. 
\scr{Out of the total 20 runA mocks from \citet{White11}, only one mock has its full initial conditions in Fourier space stored. We use this mock to calculate propagators.}
Despite our using only one simulation, the sample variance is substantially reduced because the propagator calculation involves a division of the final density field with the initial density field and both fields are subject to the same random seed. The remaining fluctuations due to a fine $k$ binning are reduced by applying Savitzky-Golay smoothing filtering \citep{SaGo}. For testing the BAO fitting models, we use all 20 mock simulations.

\section{Modeling the BAO information in the measured power spectrum}\label{sec:modelBAO}
\subsection{Estimating the signal}
We make a quantitative estimate of the BAO signal, which we will later utilize to predict the precision of the BAO measurements. 
We can directly measure the BAO signal by deriving the cross-correlation between the initial, linear density field and the final density field, often dubbed as the ``propagator"\citep[e.g.,][]{Crocce06b,Crocce08,Taruya08,Taruya09}. For biased tracres in redshift space, we normalize the propagator to be:
\begin{equation}\label{eq:propab}
C(k,\mu,z_o)=\frac{1}{b\Plin(k,z_o)}\llan\hdelta_{\rm lin}(\kvec,z_o)\hdelta^*(\kvec',z_o)\rran,
\end{equation}
where $\hdelta_{\rm lin}(\kvec,z_o)$ and $\Plin(k,z_o)$ are, respectively, the linear theory density field and the corresponding power spectrum in real space that are scaled to $z_o=0.55$  using the linear growth function; $\hdelta(\kvec,z_o)$ is the observed galaxy density field or any density field of interest at $z_o=0.55$ in Fourier space before or after BAO reconstruction. The normalization of $C(k,\mu)$ is such that it would converge to unity in real space and to $(1+\beta\mu^2)$  in redshift-space when $\hdelta(\kvec,z_o)$ contains the same information as the initial, linear density field at low  $\bk$. In linear theory the propagator is equal to $(1+\beta\mu^2)$. As the BAO feature is progressively damped during nonlinear structure growth and redshift-space distortions, $C(k,\mu)$ gradually decrease with increasing $k$.  

\subsection{Modeling the propagator}\label{subsec:prop}
The propagator estimates the BAO damping as a function of wavenumber and therefore has been useful for calculating the expected precisions of a BAO survey as well as constructing a reasonable fitting model for the observed BAO measurement. A Gaussian damping model in Fourier space (equivalently, a Gaussian convolution for the correlation functions in configuration space) has been shown to be a fairly reasonable description of the propagator for the spherically averaged power spectrum before and after reconstruction and in real and redshift space \citep[e.g.,][]{Seo10} and indeed has been widely used in BAO fitting models when analyzing observational data.  

The production of galaxy catalogues covering ever greater cosmic volume enabled the two-dimensional standard ruler test that measures the BAO scale both along and across the line of sight. Note that the test of propagators in redshift space after BAO reconstruction conducted in \citet{Seo10} is not only for a spherically averaged power spectrum, but also assuming the anisotropic BAO reconstruction convention (`Rec-Ani'). Meanwhile, the default reconstruction used within the BOSS collaboration is the isotropic BAO reconstruction convention (`Rec-Iso', e.g., \citealt{DR11}).  We first study the dependence of the propagator on such RSD conventions and then also study their dependence on the smoothing scale. We examine the validity of the Gaussian BAO damping model in two dimensions, i.e., along the line of sight and across the line of sight. As a reminder, the Gaussian BAO damping model for a propagator is written as;

\begin{equation}
\Cg(k,\mu,z)=(1+\beta\mu^2) \exp\left[{-\frac{k^2(1-\mu^2)\Sigxy^2}{4}-\frac{k^2\mu^2\Sigz^2}{4}}\right],\label{eq:CkG}
\end{equation}
where $\Sigxy$ and $\Sigz$ represent the characteristic BAO damping scales across the line of sight and along the line of sight, respectively.

As will be probed later, we find that the two-dimensional Gaussian BAO damping model is no longer a good model for the isotropically reconstructed field, especially when using a large smoothing length during density field reconstruction. We test the following, {\it modified} Gaussian BAO damping model for the 
isotropic reconstruction (Ross et al. in preparation). 

\begin{eqnarray}
&&\Cmg(k,\mu,z)=\nonumber\\
&&(1+\beta\mu^2(1-S(k))) \exp\left[{-\frac{k^2(1-\mu^2)\Sigxy^2}{4}-\frac{k^2\mu^2\Sigz^2}{4}}\right].\nonumber\\
\label{eq:CkGmod}
\end{eqnarray}

In Appendix~\ref{sec:appa} (in Eq.~\ref{eq:CkGmodLPT}), we show that this form is a close approximation to the first order (tree-level) LPT formalism based on \citet{PadLPT, Noh2009,Mat08a,Mat08b}. The corresponding, first order  derivation for the reconstructed BAO in configuration space is presented in \citet{White15}. To be exact,  \citet{White15} presents the full Zeldovich approximation rather than the tree-level derivations for the reconstructed correlation functions for the two BAO reconstruction conventions we discuss in this paper.   In Appendix~\ref{sec:appc}, we also show the second order (one-loop level) LPT derivations for the reconstructed BAO feature in power spectra {\it in redshift space} that extends the real-space results in \citet{Noh2009} and find that the densify field before reconstruction is better described by second order LPT, while the density field after reconstruction seems to be follow first order LPT. We suspect that this happens partly because the assumptions during the BAO reconstruction such as the smoothed nonlinear density field being treated approximately as the smoothed linear density field breaks down when we consider higher order terms and/or because we truncate the second order terms at one-loop level. 

\begin{figure*}
\centering
\includegraphics[width=0.9\linewidth]{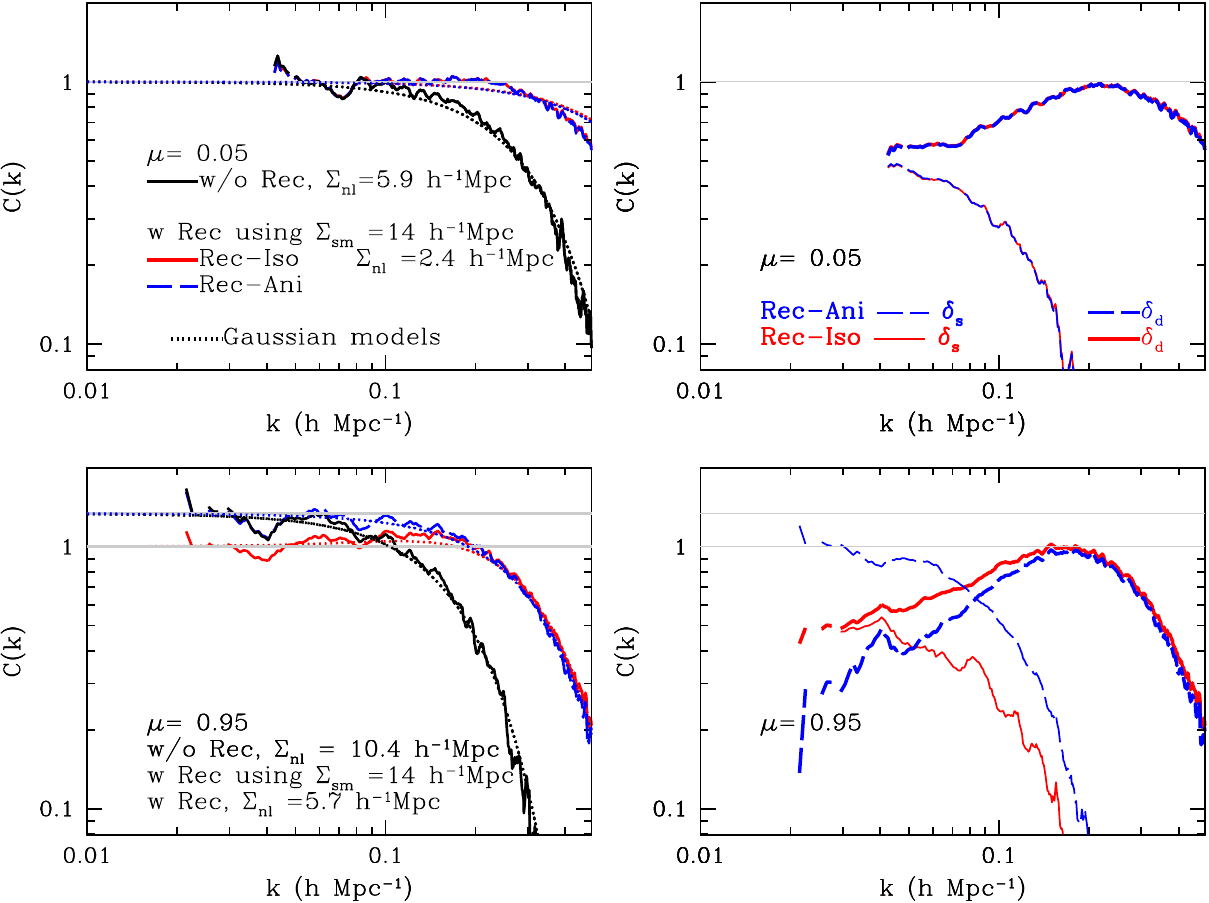}
\caption{Propagators of mock CMASS galaxy density field using different conventions for how RSD-induced anisotropy is treated in the reconstruction, for a fixed smoothing length $\Sigsm = 14\hMpc$. The black line shows the pre-reconstructed density field; the red lines show the post-reconstructed field with the isotropic reconstruction (corresponds to the magenta lines in Figure \ref{fig:figSm}) and blue \scr{dashed} lines show the anisotropic reconstruction. The left panels show propagators for $\delta_{\rm rec}$ and the right panels show the individual density fields, i.e., the negative of the displaced random field ($-\delta_s$, the two thinner curves peaking at low k) and displaced galaxy density field ($\delta_d$, the two thicker curves peaking at high k) that forms the reconstructed density field (i.e., in the left-hand panels) after mutual addition. Dotted lines show the Gaussian damping model with $\Sigma_{\rm nl}(k=0.3\ihMpc)$: the red dotted line shows the modified Gaussian damping model while the blue dotted line shows the original Gaussian damping model. }
\label{fig:figAni}
\end{figure*}

\subsection{Estimating the noise}
We consider the nonlinear power spectrum to be the sum of the signal term $C(k,\mu,z)^2 \Plin (k,z)$ that contains the BAO information and the rest of the term that does not directly relate to the BAO information. The latter is often called the mode-coupling term \citep{Crocce08,Mat08a} and we use $\Pmc$ to represent it:
\begin{equation}
\Pnl(k,\mu,z_o)=C(k,\mu,z_o)^2 \Plin (k,z_o) + \Pmc (k,\mu,z_o).
\end{equation}\label{eq:Pnl}
The mode-coupling term includes shot noise as well as a very weak, phase-shifted BAO remnant which introduces a small systematic bias to the BAO measurement depending on the relative amplitude of this term. The systematic bias is effectively removed by BAO reconstruction \citep[e.g.,][]{Seo08,PadCal} and we will ignore this aspect of $\Pmc$ in this paper.  Instead, we are interested in quantifying the noise contribution from $\Pmc$.

The measurement noise of a power spectrum is proportional to the amplitude of the measured power spectrum itself, $\Pnl$, ignoring the tri-spectrum contribution. When the signal we want to measure is the BAO information, \ie, $C(k,\mu,z)^2 \Plin (k,z)$, a larger $\Pmc$ therefore means a larger noise level.
In summary, the signal-to-noise ratio of the BAO measurement with the post-reconstructed power spectrum would increase as the damping effect in $C(k,\mu,z)$  decreases (i.e., as approaches to $1+\beta\mu^2$) and as the amplitude of $\Pmc$ decreases.
In the next section, we investigate how $C(k,\mu)$ and $\Pmc$ change depending on which BAO reconstruction convention is used.  We will eventually incorporate these results into the Fisher matrix analysis and predict the precision of BAO distance measurements.

\section{Results: The BAO signal} \label{sec:results-signal} 
In this section, we use the runA mock simulations introduced in \S~\ref{sec:mock} to study the propagator (as an indicator for the BAO signal strength) and $\Pmc$ (as noise) before and after reconstruction as a function of RSD convention and smoothing scale.
The dependence of the reconstructed BAO on these various details has been investigated in the literature, but in a way that is subject to a large sample variance, i.e., in terms of means and standard deviations of the final BAO constraints from mock simulations \citep{Pad12,Burden14,Vargas15}. Such analysis requires a very large ensemble of simulations to reduce the sample variance on the error estimates and could be sensitive to the details in the fitting procedure.  On the other hand, calculating propagators does not require a large number of mocks and presents the BAO signal as a function of scale in a way that can be directly translated to Fisher matrix calculations. The propagator has also proven to be useful for the construction of models to extract BAO information from clustering measurements.

\subsection{Varying the anisotropy convention}

\begin{figure*}
 \centering
 \includegraphics[width=0.9\linewidth]{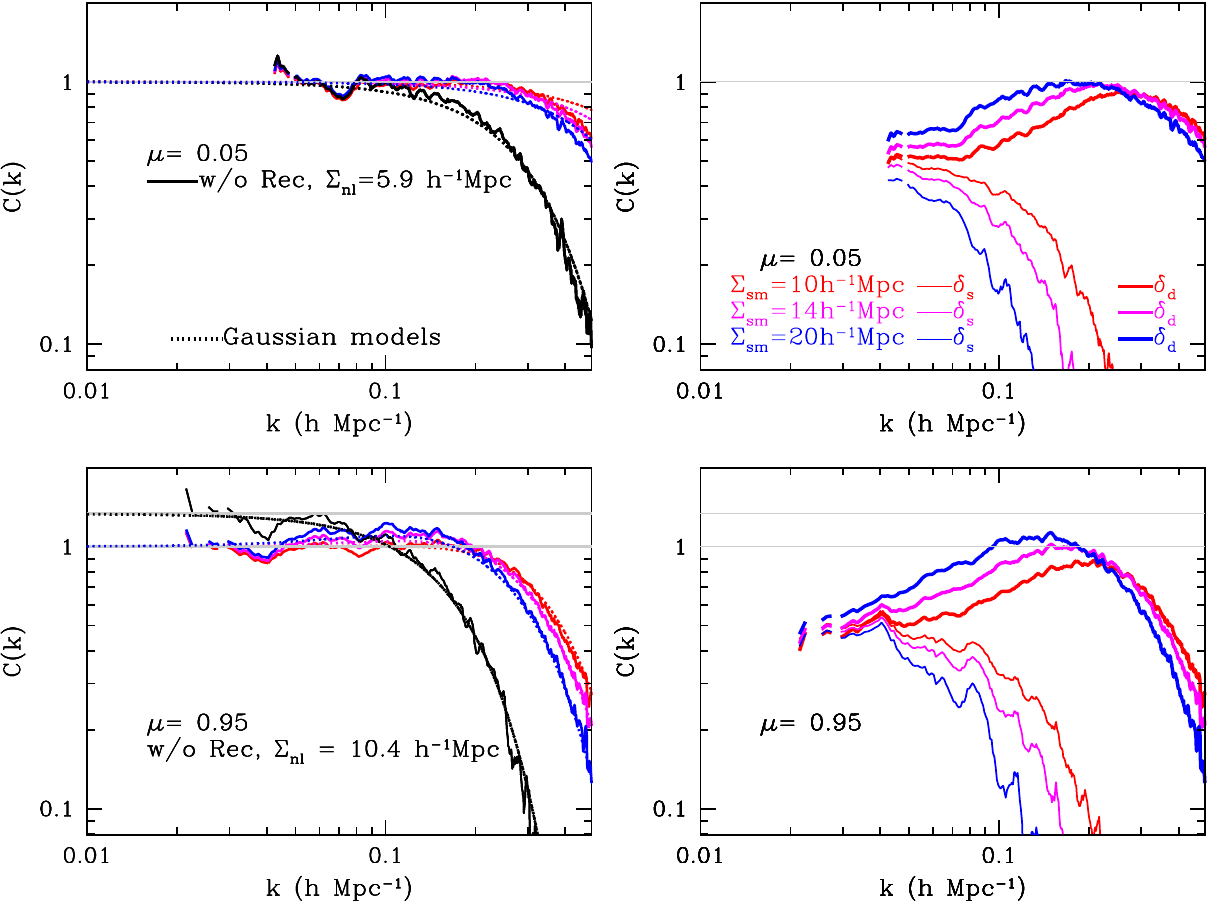}
 \caption{Left-hand panels: propagators of the mock CMASS galaxy density field before (black) and 
after reconstruction as a function of smoothing length $\Sigsm$: red, magenta, and blue correspond to $\Sigsm = 10, 14, 20\hMpc$. The right-hand panels show the corresponding individual density fields, i.e., the displaced random field  ($-\delta_s$, the set of three thinner curves peaking at low $k$) and 
displaced galaxy density field ($\delta_d$, the set of three thicker curves peaking at high $k$) that forms the reconstructed density 
field (i.e., in the left-hand panels) after mutual addition. Top panels show the propagators for the modes across the line-of-sight direction (cosine the line-of-sight angle $\mu=0.05-1$) and the bottom panels show the propagators for the modes almost along the line of sight ($\mu = 0.95-1$). Black dotted lines show the Gaussian damping model with $\Signl(k=0.3\ihMpc)$ before reconstruction. Red, magenta, and blue dotted lines show the modified Gaussian damping model with $\Signl(k=0.3\ihMpc)$. Gray reference lines are located at unity and $(1+\beta\mu^2)$. }
\label{fig:figSm}
\end{figure*}

\begin{figure}
\centering
\includegraphics[width=0.9\linewidth]{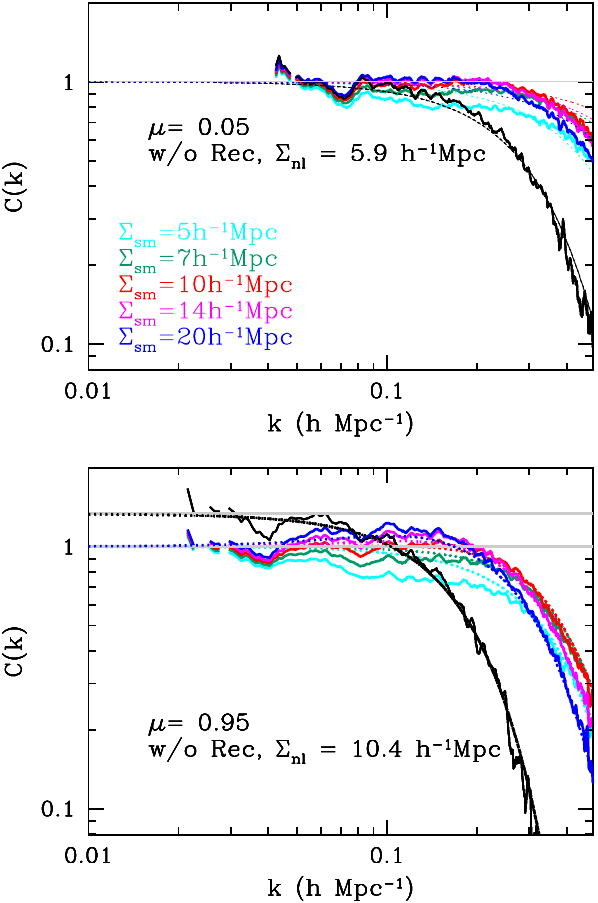}
\caption{The same as the left panels of Figure \ref{fig:figSm} except that cyan and green lines correspond to $\Sigma=5$ and $7~\hMpc$. Red, magenta, and blue correspond to $\Sigsm = 10, 14$, and $20\hMpc$ as before. }\label{fig:figSmt}
\end{figure}

We first compare propagators for the two different RSD-induced anisotropy conventions discussed in \S~\ref{subsec:anirec}: the original, anisotropic reconstruction (Rec-Ani, Eq. \ref{eq:recani}) and the isotropic reconstruction (Rec-Iso, Eq. \ref{eq:reciso}). As explained earlier, both methods differ in several details, but the main difference is whether or not to restore the linear redshift-space distortions in displacing the random particles.

The left panel of Figure~\ref{fig:figAni} shows the propagator of the mock CMASS galaxy density field before (black) and 
after reconstruction (red lines for Rec-Iso and blue \scr{dashed} lines for Rec-Ani) with $\Sigsm = 14\hMpc$\scr{, which is an intermediate smoothing scale between our estimated optimal smoothing scale of $10\hMpc$ (in \S~\ref{sec:con}) and $20\hMpc$ that has been used in previous BOSS analysis~\citep{DR11}. } 
The top panel shows modes across the line of sight (defined by $\mu=0-0.05$) and the bottom panel shows modes along the line of sight (defined by $\mu = 0.95-1$). As expected, the propagator for the pre-reconstructed field in black, asymptotically approach\scr{es} unity at small $k$ for the modes across the line of sight and approaches $(1+\beta\mu^2)$ for the modes along the line of sight. The propagator for the post-reconstructed field should converge to unity at small $k$ in both directions in the isotropic reconstruction convention, while they should converge to $1+\beta \mu^2$ in the anisotropic convention. The figures in the left panel indeed show such a behavior. 
For the isotropic reconstruction case, it is evident that the propagator overshoots above unity in the intermediate scales along the line of sight, which is consistent with the effect of $1-S(k)$ in Eq. \ref{eq:CkGmod}. 
\scr{From the line-of-sight propagator curves in the lower left panel, one can see that $C(k=0.3\ihMpc)/C(k=0\ihMpc)$ will be greater in the case of the isotropic reconstruction (i.e., $(1+\beta(1-S(k)))\exp(-\Signl^2 k^2/4)|_{k=0.3\ihMpc}$) in comparison to the anisotropic case (i.e., $\exp(-\Signl^2 k^2/4)|_{k=0.3\ihMpc}$), which implies that the restored BAO feature from the isotropic case would correspond to smaller effective damping compared to the anisotropic reconstruction case albeit the very similar $\Signl$ values in both cases. }
The question of which convention would be more efficient in terms of the BAO signal to noise would also depend on the noise component in each case, which will be investigated in \S~\ref{sec:results-noise}.

Right panels show the corresponding propagators for the displaced random field alone (to be exact, the negative of the displaced random field, -$\delta_s$, in Eq ~\ref{eq:drec}; the blue and red thinner curves peaking at low $k$) and the propagators for
the displaced galaxy density field alone ($\delta_d$ in Eq \ref{eq:drec}; the blue and red thicker curves peaking at high $k$) that together comprise the reconstructed density field shown in the left panels ($\delta_{\rm rec} = \delta_d-\delta_s$, Eq. \ref{eq:drec}). We find that the difference in damping is mainly caused by the contribution from the displaced random field $\delta_s$ being suppressed relative to the contribution from the displaced mock galaxy field $\delta_d$ in the isotropic reconstruction. Since the contribution from $\delta_d$ is very similar in the two conventions at large $k$, so is the propagator of the final field $\delta_{\rm rec}$ at large $k$. Across the line of sight (top panels), both conventions produce almost identical propagators, which is expected as RSD should not affect these modes.

In the left-hand panels of Figure \ref{fig:figAni}, we over-plot the corresponding Gaussian damping models (Eq. \ref{eq:CkG} for black and blue and Eq. \ref{eq:CkGmod} for red lines) in dotted lines that are estimated based on the propagator values at  $k=0.3\ihMpc$; our choice of $k=0.3\ihMpc$ is made based on the visual inspections~\footnote{\scr{I.e., it is done by visually inspecting the overall agreement between the Gaussian model anchored at different choices of $k$ values and the measured propagator, rather than assuming errors on the measured propagators and thereby deriving the best fit Gaussian models.}}.  Note that in both directions the Gaussian damping model describes the propagator fairly well for the pre-reconstructed field (black lines).  $\Signl$ values labeled in the figure correspond to the damping scales for the dotted lines and we interpret these values at $\mu=0-0.05$ (top panels) and $\mu=0.95-1$ (bottom panels) to be $\Sigxy$ and $\Sigz$ in Eq \ref{eq:CkG} (or \ref{eq:CkGmod}), respectively.  

We find that the propagators after the anisotropic reconstruction are well modeled by the default Gaussian damping model, while the isotropic reconstruction is better modeled by the modified Gaussian model, when focusing on the line-of-sight direction ($\mu \sim 1$). 
\scr{Despite the different forms, the derived $\Signl$ values at $\mu\sim 1$ are almost identical in the two cases. For example, the measured $\Sigz$ values as shown in Table~\ref{tab:fisher} are $5.6\hMpc$ and $5.7\hMpc$ in the case of isotropic and the anisotropic case, respectively, when reconstructed with $\Sigsm=14\hMpc$. This implies the approximate forms from LPT in the Appendix (Eq.~\ref{eq:CkGLPT} and Eq.~\ref{eq:CkGmodLPT} ) work well. Again note that the $1-S(k)$ in Eq. \ref{eq:CkGmod} makes the {\it effective} BAO damping after reconstruction smaller in the isotropic case albeit almost the same $\Signl$.}

Across the line of sight, \ie, when $\mu \sim 0$, the modified Gaussian damping model of course asymptotically becomes the Gaussian damping model. The red solid and dotted lines in the top left panel apparently show that such Gaussian damping model is a poor description on small scales. We nevertheless note that the model reproduces the near unity convergence for $k \le 0.3\ihMpc$ which is the wavenumber range we use in the BAO analysis and therefore do not further modify the model to improve its behavior at $\mu\sim 0$ when calculating Fisher matrix. 

In summary, we showed that the anisotropic reconstruction indeed has a boosted signal normalization on large scales along the line of sight, while the isotropic reconstruction convention has an advantage of smaller effective damping along the line of sight. The smaller damping in the isotropic case, however, is caused by a relative suppression of the displaced random particles on large scales.  That is, the curve of propagator $C(k)$ on smaller scales (caused by the displaced  galaxies) is almost identical in both cases.
If we assume the same level of noise for the post-reconstructed density field in both reconstruction conventions, this would imply more BAO information in the anisotropic case, despite the shallower damping in the isotropic reconstruction. However, we expect that not only the signal but also the noise would have been boosted by the same factor in the anisotropic case. In \S~\ref{sec:results-noise}, we therefore check the noise level of the reconstructed density field in both conventions. As a caveat, \citet{DR11} used the default Gaussian damping model when fitting for the BAO feature that was reconstructed using the isotropic reconstruction.
 Ross et al. (in prep) and Beutler et al. (in prep) will utilize the modified Gaussian damping model, and we expect these results will be included in the final BOSS analyses (Anderson et al. in prep.); this paper demonstrates the necessity for such update.

\subsection{Dependence on the smoothing term}\label{subsec:signalsm}
The effect of the smoothing term $S(k)$ used to filter the nonlinear density field during reconstruction has been investigated in the literature \scr{\citep[e.g.,][]{PadLPT,Seo10, Mehta11,Vargas14,Burden14,Vargas15,Achitouv15}}. We attempt to summarize those results in what follows. First, it is expected that the effect would depend on the extent of nonlinearity (by affecting the validity of Eq.~\ref{eq:qdiv}) and shot noise. If our goal is optimally deriving the true $S(k)\hat{\delta}(k)$ where $\hat{\delta}(k)$ is the true Fourier-space density field, $S(k)$ should be weighted with $1/[1+nP(k)]$ (e.g., \citet{SeoHirata15}). 
\citet{White10} analytically derived the expected efficacy of real-space BAO reconstruction (in terms of the expected Gaussian damping of the propagator) as a function of the smoothing length and shot noise within the context of LPT; the paper predicts that the efficacy will improve as shot noise decreases while such improvement starts to saturate around $\nPt \sim 0.2$-$1$.\footnote{\citet{White10} quotes $\nbar = 10^{-4}\trihMpc$ for $b=1$ for the saturation point. Assuming $P(k=0.2) \sim 2000 \trihMpc$, this gives $\nPt \sim 0.2$.
 }
 As a function of the smoothing scale $\Sigsm$, \citet{White10} predicts that the efficacy will be optimal over a broad range of $\Sigsm$, while the optimal range shifts to a larger smoothing length for a noisier field.  
In redshift space, this real-space prediction could be applied to approximately predict the behavior for the modes across the line of sight.

In this subsection, we focus on the isotropic reconstruction convention and test the reconstruction efficacy in redshift space as a function of the smoothing scale \footnote{Some examples of the anisotropic reconstruction are presented in \citet{Seo10}.}. 
We consider the signal to noise level of BOSS CMASS.

\begin{figure*}
\centering
\includegraphics[width=0.9\linewidth]{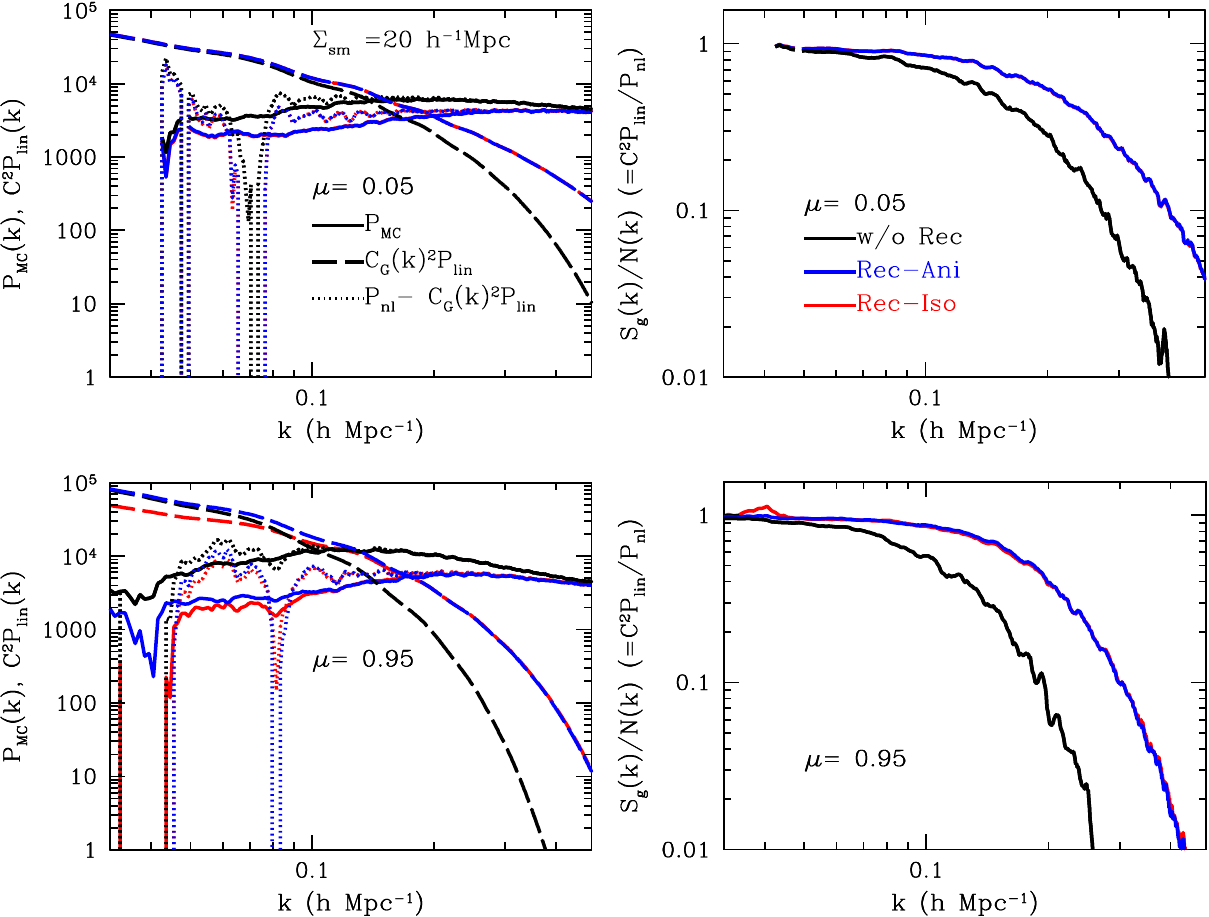}
\caption{
Left: Mode-coupling components (solid lines, $\Pmc=\Pnl-C(k,\mu)^2\Plin$) relative the BAO signal estimated by $C_G(k,\mu)^2\Plin$ (dashed lines) for the two anisotropy conventions while fixing  smoothing length $\Sigsm = 20\hMpc$. Black line: pre-reconstructed density field. Red: post-reconstructed field with the isotropic reconstruction. Blue: using the anisotropic reconstruction. \scr{In the upper left and right panels for $\mu=0.05$, the blue line and the red line are almost over-imposed.} The dotted black lines show $\Pmc$ when using the corresponding Gaussian models for the propagators, i.e., $\Pmc=\Pnl-\Cg(k,\mu)^2\Plin$ for the pre-reconstructed field and $\Pnl-\Cmg(k,\mu)^2\Plin$ for the post-reconstructed field.
Right: the BAO signal to noise ratio for each $k$ mode, i.e.,  $C^2\Plin/\Pnl$. Note that, with $\Pnl$ in the denominator, the `noise'  is now the total noise that includes the sample variance from the signal $C(k,\mu)^2\Plin$ in additional to $\Pmc$.
}
\label{fig:figAniPk}
\end{figure*}

The left-hand panels of Figure~\ref{fig:figSm} show the propagators of our mock CMASS galaxy density field before (black) and 
after (colored lines) the isotropic reconstruction as a function of smoothing length $\Sigsm$: red, magenta, and blue correspond to $\Sigsm = 10, 14$, and $20\hMpc$, respectively. 
The figure shows that the propagators for the post-reconstructed field depend on the smoothing length and that the behaviors in the two directions are different.  For the modes across the line of sight (top panel), they all asymptotically approach unity on large scales (small $k$) while a smaller smoothing length causes a convergence at slighter larger $k$, i.e., a more efficient reconstruction. 
For the modes along the line of sight (bottom panel), as evident in the blue line with $\Sigsm=20\hMpc$ relative to the red line with $\Sigsm=10\hMpc$, the overshooting above unity in the intermediate scales becomes more severe for a larger $\Sigsm$; the original Gaussian damping model (Eq.~\ref{eq:CkG}) is a worse description for a larger $\Sigsm$.  
Red, magenta, blue dotted lines show the corresponding modified Gaussian damping model (Eq. \ref{eq:CkGmod}) for the post-reconstructed density field; we derive $\Sigxy=2.0, 2.4$, and $2.9\hMpc$ and $\Sigz=5.1, 5.6$, and $6.2\hMpc$ (in Eq.~\ref{eq:CkGmod}) for them, respectively, showing a smaller BAO damping scale when reconstructed with a smaller smoothing scale.
That is, when $\Sigsm$ changes from $20\hMpc$ to $10\hMpc$, the decrease in the BAO damping scale is 45\% across the line of sight, while it is 21\% along the line of sight;  decreasing the smoothing scale is less effective along the line of sight. 

The right-hand panels of Figure~\ref{fig:figSm} show the individual contributions from $\delta_d$ and $-\delta_s$. Using a smaller smoothing length (e.g., the red line rather than the blue line) increases the contribution from $\delta_s$, since the reconstructed displacement field has more information extending to smaller scales, and shifts the contribution from $\delta_d$ to smaller scales.  In the lower right panel, note the overshoot above unity by the blue $\delta_d$ line . 

When we further decrease the smoothing scale, the reconstruction efficiency saturates between $\Sigsm=7\hMpc - 10\hMpc$ and decreases when $\Sigsm < 7\hMpc$ as shown in Figure \ref{fig:figSmt}. This is reasonable as a smaller $\Sigsm$ will allow information from smaller scales that would be increasingly dominated by noise and nonlinearity.  Note that this convergence is based on the shot noise level of BOSS CMASS; dependence on $\Sigsm$ would change as a function of the shot noise level and the degree of nonlinearity. Accordingly, we find that the modified Gaussian damping model fares worse as the smoothing length decreases below $10\hMpc$, probably due to using more nonlinear and shot-noise dominated information when reconstructing the displacement field.
\section{Results: Noise} \label{sec:results-noise}
Assuming a Gaussian error, the noise level of the reconstructed field can be approximated by estimating the amplitude of $\Pmc$ after reconstruction, i.e., the component of the power spectrum that does not directly contain the BAO information \footnote{\citet{Ngan2012} has shown that non-Gaussian errors on the BAO measurement in multi-parameter fitting are negligible.}. In this section, we observe the mode-coupling terms for different smoothing scales and anisotropy conventions. Note that $\Pmc$ includes Poisson shot noise contribution.
\subsection{Varying anisotropy convention}
Figure \ref{fig:figAniPk} compares the isotropic BAO reconstruction (red lines) and the anisotropic BAO reconstruction (blue lines) for $\Sigsm=20\hMpc$.  The black lines corresponds to the pre-reconstructed density field. In the left panels, the dashed lines show $C_G(k,\mu)^2\Plin$ (BAO signal model) and the solid lines show $\Pmc=\Pnl-C(k,\mu)^2\Plin$ (noise) where $\Pnl$ includes shot noise contribution. The dotted black lines show $\Pmc$ when using the corresponding Gaussian models for the propagators, i.e., $\Pmc=\Pnl-\Cg(k,\mu)^2\Plin$ or $\Pnl-\Cmg(k,\mu)^2\Plin$.  That is, the difference between the solid and the dashed lines is that the former use\scr{s} the measured $C(k,\mu)$ while the latter use\scr{s} the Gaussian model propagators. We will adopt these Gaussian models (dotted lines) to approximate the BAO damping in the Fisher matrix calculations in \S~\ref{sec:discuss}.
As expected, because the two conventions conduct an almost identical operation across the line of sight, we observe no difference in $\Pmc$ in this direction (\scr{i.e., blue and red lines are over-imposed in the top left panel}).
Meanwhile, $\Pmc$ from the two conventions are almost identical also along the line of sight (bottom left panel) for $k > 0.1\ihMpc$ while being slightly different for $k < 0.1\hMpc$. This range of $k <  0.1\ihMpc$ is approximately where the signal terms of the two conventions are different along the line of sight because each convention makes a different choice on displacing random particles. Note that $\Pmc$ is slightly higher for the anisotropic reconstruction case along with the stronger BAO signal for this convention. This is reasonable; on large scales, the dominant contribution to $\Pmc$ would be shot noise and any shot noise relative to the signal in the density field we use to derive the displacement field will be frozen despite the overall factor we multiply to correct for the redshift-space distortions.   

The right panel shows the BAO signal to noise ratio for each $k$ mode, i.e., $C^2\Plin/\Pnl$ for the two different conventions; note that the noise includes the sample variance from the signal in this plot.
The two conventions return almost the same signal to noise in both directions. It is partly because, the range of $k <  0.1\ihMpc$ where the signal terms of the two conventions differ is the range where the sample variance from the cosmic signal dominates $\Pmc$, making the boosted signal ineffective, and also partly because $\Pmc$ is also slightly higher for Rec-Ani along with the stronger BAO signal. 
If we decrease to $\Sigsm=14~\hMpc$, we find that the two convention returns even more similar signal to noise ratios in both directions. This result agrees with \citet{Burden14} that found no difference in the BAO precisions between the two conventions at the noise level of BOSS CMASS sample.
\begin{figure*}
 \centering
 \includegraphics[width=0.9\linewidth]{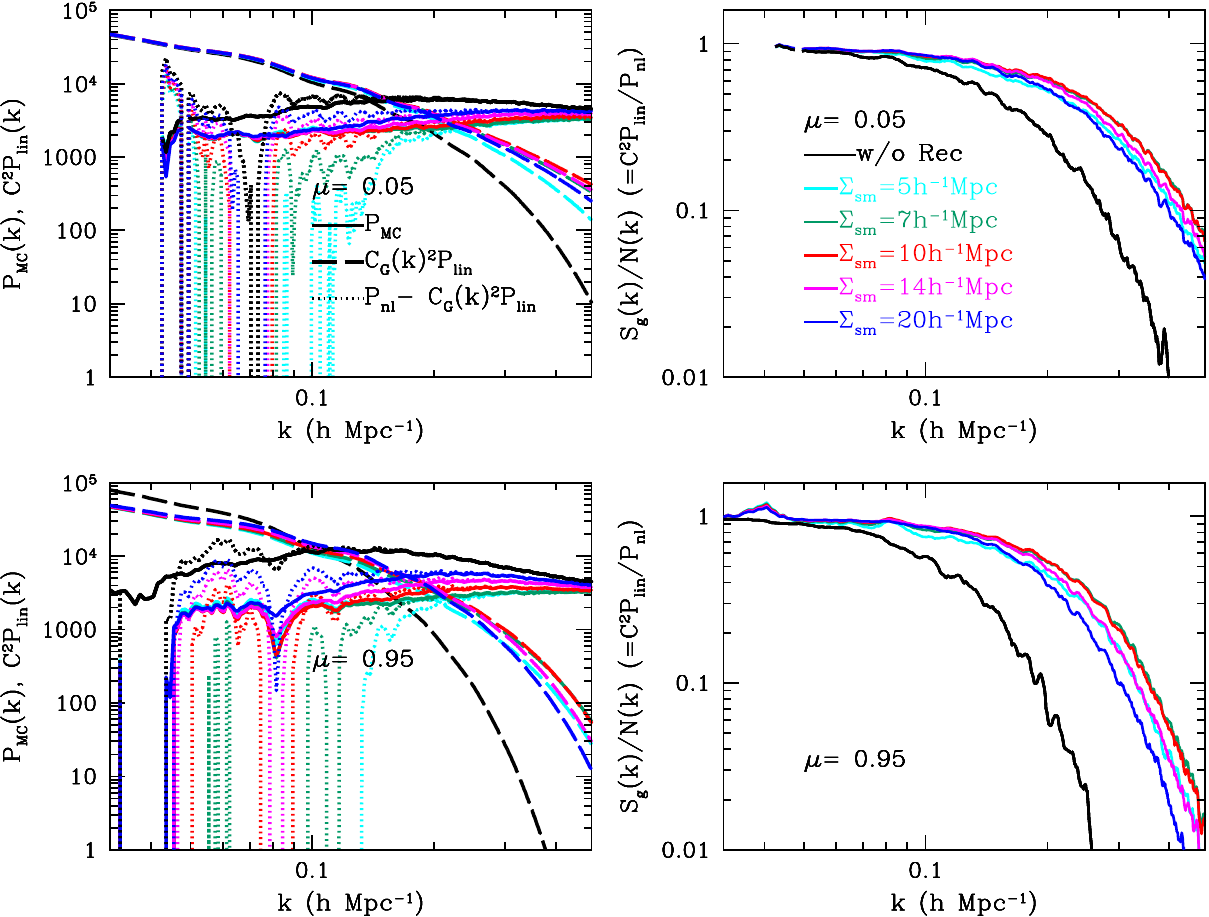}
\caption{Left: Mode-coupling components (solid lines, $\Pmc=\Pnl-C(k,\mu)^2\Plin$) relative the BAO signal estimated by $C_G(k,\mu)^2\Plin$ (dashed lines). Cyan, green, red, magenta, and blue correspond to the reconstructed density field using $\Sigsm = 5, 7, 10, 14$, and $20\hMpc$ and the black corresponds to the pre-reconstructed density field. The dotted black lines show $\Pmc$ when using the corresponding Gaussian models for the propagators, i.e., $\Pmc=\Pnl-\Cg(k,\mu)^2\Plin$ for the pre-reconstructed field and $\Pnl-\Cmg(k,\mu)^2\Plin$ for the post-reconstructed field. Right: the BAO signal to noise ratio for each $k$ mode, i.e.,  $C^2\Plin/\Pnl$. Note that, with $\Pnl$ in the denominator, the `noise'  is now the total noise that includes the sample variance from the signal $C(k,\mu)^2\Plin$ in additional to $\Pmc$.  \scr{In the top right panel, the green ($\Sigsm=7\hMpc$) and the red ($10\hMpc$) lines are almost super-imposed while the cyan ($5\hMpc$) and magenta ($14\hMpc$) are almost super-imposed in the bottom panel.}  
} \label{fig:figSmPk}
\end{figure*}

\subsection{Dependence on the smoothing term}
In Figure \ref{fig:figSmPk}, red, magenta, and blue correspond to the reconstructed density field using $\Sigsm = 10, 14$, and $20\hMpc$, respectively.  Cyan and green correspond to $\Sigsm = 5\hMpc$ and $7\hMpc$.
The left panel shows that as the smoothing length decreases from $20\hMpc$ to $10\hMpc$ (from blue to red), $\Pmc$ of the post-reconstructed field further decreases in both directions, but more so along the line of sight.  In other words, decreasing the smoothing length not only improves the BAO signal (based on the propagator) but also decreases the nonlinear effect on small scales. Such a decrease in $\Pmc$ tends to saturate below $\Sigsm \sim 7\hMpc$ (i.e., the cyan solid line overlaps with the green solid line) possibly due to the domination of shot noise and nonlinearity on these scales or due to the resolution of the mesh size (i.e., $2.9\hMpc$) used to measure the power spectrum in the mock simulation.   

In the case of $\Sigsm = 20\hMpc$ (blue), we find that $\Pmc$ (left panels) derived using the modified Gaussian damping model (dotted lines) is somewhat greater than the exact estimate on very large scales (solid lines); note, however, that the signal dominates noise on this scale. In the case of  $\Sigsm = 5\hMpc$ (cyan), $\Pmc$ using the modified Gaussian damping model is a poor description, as also implied in Figure~\ref{fig:figSmt}. We therefore focus on $\Sigsm = 10, 14$, and $20\hMpc$ when conducting the Fisher matrix analysis in \S~\ref{sec:discuss} using the Gaussian damping models. 

Traditionally, we use the signal to noise of power at $k=0.2\ihMpc$ to estimate the BAO constraints \citep{SE07}. We derive $\Pmc= \Pnl-\Cg^2\Plin$ at $k=0.2\ihMpc$, i.e., $\Pmct$ after reconstruction, which returns 2604, 3066, and $3697~\trihMpc$ for the modes across the line of sight and 3413, 4472, $5819~\trihMpc$ along the line of sight for $\Sigsm = 10, 14$, and $20\hMpc$, respectively. The anisotropy in $\Pmct$ is greater for a larger smoothing length: the ratio of $\Pmct$ between the two directions is 1.31 when $\Sigsm=10\hMpc$ while it is 1.57 when $\Sigsm = 20\hMpc$.

The right panel clearly shows the well-known significant signal-to-noise gain due to reconstruction, which increases with decreasing smoothing length. The gain becomes saturated when $\Sigsm=7-10\hMpc$ and the signal to noise reduces as the smoothing length is reduced below $7\hMpc$. \citet{Vargas15} studied the dependence on the smoothing length by estimating dispersions of the BAO constraints among mock simulations that are reconstructed with the isotropic reconstruction. They are using the original Gaussian damping model for the BAO fitting, while we are relying on a pure Fisher matrix approach. Their result agrees with ours in terms of the optimal smoothing lengths after noting that their derived optimal smoothing lengths of $5-10\hMpc$ correspond to $7-14\hMpc$ in our convention. 

Taking the validity of the modified model into consideration shown in the previous section, the smoothing scale of $10\hMpc$ appears to be the optimal choice for the BAO analysis. 
These plots also support the robustness of the reconstruction efficiency despite the choice of smoothing length, i.e., the large difference between the black line and any of the colored lines, which again agrees with what is observed in the literature \citep[e.g.,][]{Mehta11}. 

\section{Discussions}\label{sec:discuss}
\begin{table*}
\centering
\caption{\label{tab:fisher} Fisher matrix estimates for the BOSS DR12 CMASS BAO constraints for various cases. We assume the effective survey volume of $2.36~\trihGpc$ and $z=0.55$. `Rec-Iso' refers to the isotropic reconstruction case (i.e., removing RSD) and `Rec-Ani' refers to the anisotropic reconstruction case (i.e., keeping RSD).}
\begin{tabular}{ccccccccc}
\hline\hline
 & $\Sigsm$ & $\Sigxy~(\hMpc)$ & $\Sigz~(\hMpc)$ & $A_1 (\trihMpc)$ & $A_2 (\trihMpc) $ & $\sDA(\%)$ & $\sHz(\%)$ & $\sDV(\%)$ \\
 \hline
Pre-Rec & 0 & 5.9 & 10.4 & 1193 & 4968 & 1.50 & 2.86 & 1.05 \\
\hline
Rec-Iso & 10 &  2.0 & 5.1 & 1681 & 923 &  0.81 & 1.31 &  0.54\\
Rec-Iso & 14 & 2.4  & 5.6 & 1463 & 1603 &  0.87 & 1.44 & 0.58 \\
Rec-Iso & 20 & 2.9 & 6.2 & 1277 & 2420 & 0.96 & 1.63 & 0.65 \\
\hline
Rec-Ani & 14 & 2.4 & 5.7 & 1493 & 1573 & 0.86 & 1.38 & 0.57 \\
Rec-Ani & 20 & 2.9 & 6.2 & 1438 & 2232 & 0.95 & 1.57 & 0.63\\
\hline\hline
\end{tabular}
\end{table*}

\subsection{Fisher matrix estimates of the BAO measurement precisions}
Given the dependence of the BAO signal and noise on different reconstruction details observed in this paper, we predict the BAO distance measurement precisions for the post-reconstructed density field of BOSS DR12 data as a function of the smoothing scale and the an/isotropic conventions. We use the Fisher matrix formalism in \citet{SE07} for the BAO-only signal, but appropriately modify the Gaussian damping model and the mode-coupling amplitudes. We define the signal at $\bk$ to be
\begin{eqnarray}
\mathcal{S}g(\bk) &=& \Plin(k)C^2(k,\mu) \nonumber \\
&=& \Plin(k)(1+\beta\mu^2(1-cS(k)))^2 \times \nonumber \\
&& \exp{\left[{-\frac{k^2(1-\mu^2)\Sigxy^2}{2}-\frac{k^2\mu^2\Sigz^2}{2}}\right]}, \nonumber \\
\end{eqnarray}
where $S(k)$ is the smoothing filter from Eq.~\ref{eq:sk} and $c=1$ for the post-reconstructed field with the isotropic BAO reconstruction but $c=0$ otherwise. The square root of the variance of the signal at $\bk$ is proportional to the observed total power at $\bk$, which we model as:
\begin{eqnarray}
\mathcal{N}(\bk) & =& \Pnl = \Plin(k)C^2(k,\mu) + \Pmct(\mu),
\end{eqnarray}
where we allow $\Pmct(\mu)$ to be angle-dependent; Figure \ref{fig:figAniPk} shows a weak scale-dependence in $\Pmc(k)$, but we ignore the scale-dependence and approximate $\Pmc(k) \simeq \Pmct$.

Assuming the likelihood function of the band powers of the galaxy
power spectrum to be Gaussian, the signal to noise of the band power can be approximated as
\begin{eqnarray}
\frac{\mathcal{S}g}{\mathcal{N}}&=&\sqrt{\frac{2\pi k^2dk d\mu\Vsur}{2(2\pi)^3}}\times \nonumber\\
&&\frac{\Plin(k)\exp{\left[ -\frac{\kperp^2\Sigxy^2+\kpar^2\Sigz^2}{2}\right]}}
{\Plin(k) \exp{\left[ -\frac{\kperp^2\Sigxy^2+\kpar^2\Sigz^2}{2}\right]} + {\Pmct(\mu)\over R(\mu)}}, \nonumber\\
\end{eqnarray}
where we define $R(\mu)\equiv [1+\beta\mu^2(1-cS(k))]^2$. The only difference from \citet{SE07} is therefore what goes into ${\Pmct(\mu)/R(\mu)}$.

Following \citet{SE07}, we project this $\frac{\mathcal{S}_g}{\mathcal{N}}$ on to $\DA$ and $H$. The Fisher matrix for the BAO-only information is then,
\begin{eqnarray}
&&F_{ij}= \Vsur A_0^2
\int_{0}^{1} d\mu\; f_i(\mu) f_j(\mu) \int_{0}^{\infty} dk\, \nonumber \\
&&{k^2 \exp\left [-2(k\Sigma_s)^{1.4}\right] \exp\left[-k^2(1-\mu^2)\Sigxy^2-k^2\mu^2\Sigz^2\right] \over
\left({P(k)\over\Pt} \exp\left[-k^2(1-\mu^2)\Sigxy^2-k^2\mu^2\Sigz^2\right]+{\Pmct \over [\Pt R(\mu)]}\right)^2}\nonumber\\ 
&&\label{eq:F2D} 
\end{eqnarray}
where $\Vsur$ is the volume of the survey, $\Sigma_s$ is the  Silk damping scale, and $f_i(\mu)$ and $f_j(\mu)$ are the derivatives of the BAO peak location with respect to the anisotropic distances such as angular diameter distance $\DAA$ and the Hubble parameter $\hzz$.  We take $A_0=0.4529$ and $\Sigma_s=7.76~\hMpc$ presented in \citet{SE07} for WMAP1.
For $\Sigxy$ and $\Sigz$ values, we adopt damping scales that were measured directly from the propagators at $\mu=0.05$ and 0.95.  We approximate $\Pmct(\mu)$ with $A_1+A_2(1+\beta\mu^2)^2$  while $A_1$ and $A_2$ are derived from $\Pmct$ for the modes centered at $\mu=0.05$ and 0.95.  

We find that when $\Sigsm$ decreases from $20\hMpc$ to $10\hMpc$, we expect the precision on $\DAA$, $\hzz$, and the isotropic distance scale $\DV$ \footnote{The $\DV$ constraint here is at fixed Alcock-Pazynski \citep{1979Natur.281..358A} shape $D_AH$, corresponding to a BAO distance scale derived from a spherically averaged clustering information.} to improve by $16\%$, 20\% and 17\%. When decreasing $\Sigsm=14\hMpc$ to $10\hMpc$, we predict an improvement of 10, 12, and 11\% on $\DAA$, $\hzz$, and $\DV$, respectively.

The resulting predictions for the constraints on  $\DAA$, $\hzz$, and $\DV$ for the final data release (DR12) of BOSS CMASS BAO are presented in Table \ref{tab:fisher}. The predicted constraints for DR11 CMASS data would be $\sqrt{1.15}$ times larger than these values, accounting for the volume in DR11 relative to the volume in DR12. 
For the default smoothing scale of $\Sigsm'=15\hMpc$ used for \citet{DR11}, which corresponds to $\Sigsm=20\hMpc$ in this paper,
 our prediction is $1\%$ and $0.65\%$ on $\DV$ before and after isotropic reconstruction, respectively, which is $\sim 30\%$ better than the actual DR11 constraints \citep{DR11}  even after accounting for the factor of $\sqrt{1.15}$. \scr{This discrepancy is roughly consistent with the difference between DR9 \citep{2012MNRAS.427.3435A} constraints and the Fisher predictions that \citet{DESI} pointed out. Referring to \citet{DESI}, the reason for this discrepancy could be due to Fisher matrix being overly optimistic or due to the analysis of the mocks and data being sub-optimal. Here, we test if improving the fitting model in the analysis decreases this discrepancy.  }

\subsection{Improving the fitting model for the reconstructed BAO}\label{subsec:fit}
\begin{figure*}
\centering
\includegraphics[width=1.0\linewidth]{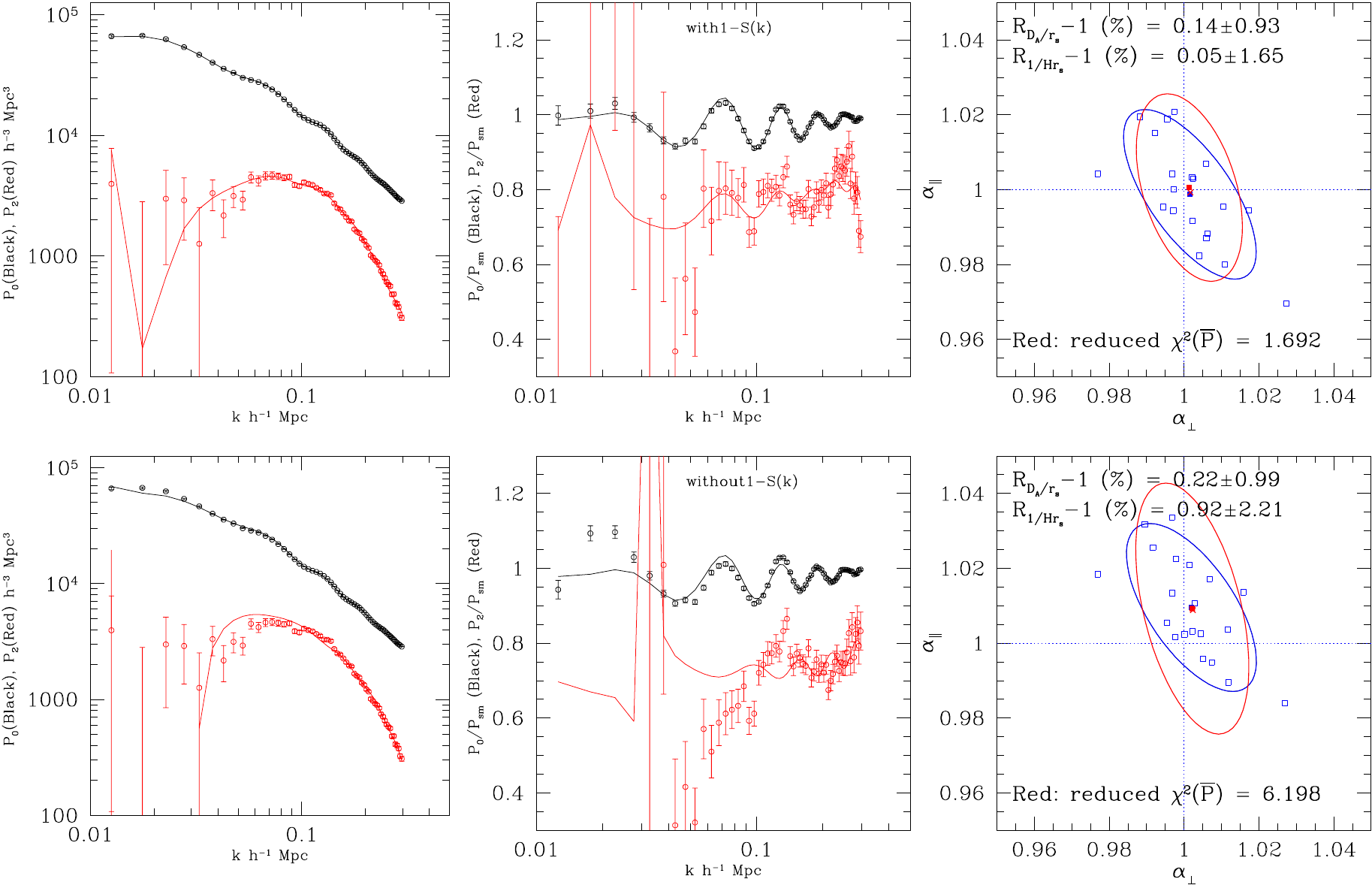}
\caption{Results of fitting power spectra multipoles using the two BAO models. The monopole (black circles) and quadrupole (red circles) power spectra are generated from 20 mock simulations that went through the isotropic reconstruction with $\Sigsm=20~\hMpc$. Shot noise is subtracted for the power spectra in this figure. The error bars correspond to the errors of the averaged power spectra and are derived from 1000 QPM mocks. Solid black (monopole) and red lines (quadrupole) show the best fits using the modified fitting model (top panels, i.e., including $1-S(k)$) and the original fitting model (bottom panels, without including $1-S(k)$). 
The middle panels show the power spectra divided by an arbitrary smooth power spectra to highlight the BAO feature. Note the large deviation between the quadrupole data and the best fit when using the original model in the bottom panel, given the errors of the averaged 20 mocks.
The right panels show the final 68\% confidence regions on  $\alpe$ and $\alpa$  (i.e., $\DA/r_s$ and $1/Hr_s$ relative to the true values); individual blue points correspond to the 20 best fits of individual mocks and the blue contour shows a covariance derived from the dispersion among the 20 best fits. Red square and cross points are the averages and the best fits of the MCMC chains using the average of the 20 mocks. The red contours are drawn based on the covariance derived from the weighted MCMC chains and rescaled to represent constraints for an individual mock.} 
\label{fig:Pmul}
\end{figure*}

Based on our results and the LPT derivations in the Appendix, we suggest using the modified Gaussian fitting model for the isotropic BAO reconstruction and the original Gaussian fitting model for the anisotropic BAO reconstruction in future surveys.  In this subsection, we test the effect of using this improved model for isotropic BAO reconstruction.
We utilize 19 more runA realizations from \citet{White11}. The left panels of Figure \ref{fig:Pmul} shows the monopole (black points) and quadrupole (red points) power spectra generated from a total of 20 mock realizations that went through {\it  the isotropic reconstruction} with $\Sigsm=20~\hMpc$. Due to the small number of runA realizations available, we derive the covariance matrix from dispersions among 1000 periodic quick particle mesh (QPM) mocks \citep{White14} that have similar monopole and quadrupole amplitudes. We rescale the covariance matrix to take into account the volume difference between the two different simulations; the error bars in the figure correspond to the rescaled errors associated with the mean of the 20 mocks.  The BAO scales, $\alpe$ and $\alpa$ are fit using a model that is consistent with Beutler et al. (in preparation):
\begin{eqnarray}
P_0(k') &=&\frac{1}{2} \int^1_{-1} P(k',\mu') d\mu' + A_0(k')\\
 P_2(k') &=&\frac{5}{2} \int^1_{-1} P(k',\mu')\mathcal{L}_2(\mu') d\mu' + A_2(k'),
\end{eqnarray}
where
\begin{eqnarray}
&&P(k',\mu') = B^2 (1+\beta \mu^2 (1-S(k)))^2\nonumber \\
&&\left[(P_{\rm lin}(k)-P_{\rm sm}(k))e^{\left[-k^2(1-\mu^2)\Sigxy^2/2-k^2\mu^2\Sigz^2/2\right]} + P_{\rm sm}(k)\right]\nonumber\\
&&\times \frac{1}{\left[1+(k\mu\Sigma_s)^2/2\right]^2}, \label{eq:Pkfit}
\end{eqnarray}
and 
\begin{eqnarray}
A_\ell(k') &=&  \frac{a_{\ell,1}}{k'^3}+  \frac{a_{\ell,2}}{k'^2}+  \frac{a_{\ell,3}}{k'}+ a_{\ell,4}+  a_{\ell,5}k'
\end{eqnarray}
Here, $k'$ and $\mu'$ are observed coordinates and they are related to the true coordinates by
\begin{eqnarray}
k&=&k'\times \frac{1}{\alpe}\sqrt{1+\mu'^2(\alpe^2/\alpa^2-1)},\\
\mu&=&\mu' \frac{1}{\alpa/\alpe\times\sqrt{1+\mu'^2(\alpe^2/\alpa^2-1)}},
\end{eqnarray}
and the no-wiggle power spectrum $P_{\rm sm}(k)$ is derived using the formula in \citet{EH98} and $P_{\rm lin}$ is derived using the same cosmology as the mock simulations.
The fitting parameters are $\alpe,\;\alpa,\;\beta,\;\ln(B^2)$, $a_{0,i}$'s, $a_{2,i}$'s, i.e., a total of 14 parameters. We interpret the final error on $\alpe$ to be a fractional error on $\DA/r_s$ and the error on $\alpa$ to be a fractional error on $1/[Hr_s]$, where $r_s$ is the sound horizon scale. We fix $\Sigxy=2.9\hMpc$, $\Sigz=6.2\hMpc$ based on the measured propagators in Table \ref{tab:fisher}; we set $\Sigma_s = 0\hMpc$ based on the consideration that the measured propagator along the line of sight should account for the finger of the God effect. \scr{Setting $\Sigma_s =1 \hMpc$ instead produces only a difference of 3-5\% compared to the default constraints we present below.}

The middle panels show the multipole power spectra divided by smooth power spectra to highlight the BAO feature. 
In the left and the middle panels, the solid black and red lines show the best fits from Monte Carlo Markov Chain (MCMC) chains using the averaged mock data, with the top panels showing the results using the modified BAO damping model (Eq., \ref{eq:Pkfit}), and the bottom panels showing the results using the original Gaussian BAO damping model (i.e., without $1-S(k)$ in Eq. \ref{eq:Pkfit}). Note that the mock data was reconstructed using the isotropic convention. 

The reconstructed field that used the isotropic convention has a small residual quadrupole but with a peak near $k\sim 0.1\hMpc$. The original Gaussian fitting model with neither zero $\beta$ nor nonzero $\beta$ can accurately describe such a peak and the imprinted BAO feature in the quadrupole as shown in the middle bottom panel. Such deviation would be less apparent when dealing with an individual mock with a larger sample variance. Indeed, on average, the reduced $\chi^2$ for an individual realization, which corresponds to 1.43 times the volume of the BOSS DR12 data, is only 1.28\;\footnote{We apply the factor from \citet{Hartlap07} to correct for the bias of the inverse covariance matrix. } using the original Gaussian model. However, when the precision of the data increases, or when we deal with a large number of mocks, which we often do to test systematics and covariance, this discrepancy becomes apparent in the form of a large reduced $\chi^2$. For the averaged power spectrum of the 20 mocks we use in this paper, we already find the reduced $\chi^2$ of \scr{6.20}
; the reduced $\chi^2$ would increase further as the sample size increased.  If we use the modified model, the reduced $\chi^2$ of the fit decreases from \scr{6.20 to 1.69} for the average of the 20 mocks (and from 1.28 to 1.04 for an individual mock). That is, the goodness of the fit has been significantly improved with the modified model.

 Another interesting question to ask is if the limitation in the current model has been limiting the BAO constraints we can achieve from the current data set so that we can use the modified model to improve the BAO constraints to the level of the Fisher predictions {\it and if inaccuracies in the model lead to biases in recovered $D_A$ or $H$ measurements}. 
The blue points of the right panels of Figure \ref{fig:Pmul} show the distribution of $\alpe$ and $\alpa$ of the best fits of the individual 20 realizations. The blue contour corresponds to the covariance constructed from the dispersions among these 20 realizations, which is subject to a large sample variance due to the small number of mocks. The red contour shows the covariance of $\alpe$ and $\alpa$ derived from weighted MCMC chains using the average power spectra of the 20 mocks. The contour has been rescaled to represent an error ellipse for an individual mock. The red cross point is the best fit from the chains and the red square is the average of the chains. All errors are corrected based on \citet{Percival14} to propagate errors in the covariance matrices derived from mocks.

We find that the modified model indeed decreases biases on the recovered  $\alpe$ and $\alpa$ measurements; the red and blue points in the top panel lie near the true cosmology, i.e. $\alpe=1$ and $\alpa=1$. Also, the consistency between the red and blue contours, i.e., between the dispersion among individual mocks and the likelihood contour of the averaged mock has been noticeably improved. 
The standard deviations from the MCMC chains give $\sDA=0.93\%$ and $\sHz = 1.65\%$ with the modified model and $\sDA=0.99\%$ and $\sHz = 2.21\%$ with the original model, which is a $\sim 30\%$ improvement in $\sHz$.

When the results for the original model are rescaled for the survey volume of DR12 (DR11), we find 1.18\% (1.27\%) and 2.64\% (2.83\%), which reasonably agrees with 1.43\% and 2.52\% from \citet{DR11} for DR11. That is, our result using the original fitting model for the isotropically reconstructed field indeed agrees with the current constraints in the literature. As pointed out in \citet{DESI}, the current DR11 constraints are about $30\%$ worse than the Fisher matrix predictions; this paper shows that the discrepancy appears to remain at this level even after considering the dependence on the reconstruction details in the Fisher matrix calculations, as investigated in Table \ref{tab:fisher}. 

When the results {\it using the modified model} are rescaled to DR12 (DR11), we find 1.11\% (1.19\%) and 1.97\% (2.11\%). Note that these values agree better with Fisher forecasts in Table \ref{tab:fisher} for the $\Sigsm=20\hMpc$ but remain 15-20\% larger.  
\citet{Ross15} showed that in the limit where the BAO information is equally distributed in each direction, i.e., in $\mu$, the hexadecapole contribution would be minimal. Under this assumption, they found a good agreement between their error forecasts and the DR11 measurements. In this paper, we find that such assumption is not strictly valid even in the isotropic reconstruction case. For example, Figure \ref{fig:figSmPk} shows  that the reconstructed BAO signal to noise ratio for each k mode is highly anisotropic, especially for a large smoothing length; even for the isotropic reconstruction case, we see the signal to noise at $k\sim 0.2\ihMpc$ is 38\% greater across the line of sight (i.e, $0 < \mu \leq 0.1$) compared to along the line of sight  (i.e, $0.9 < \mu <  1$) for $\Sigsm=20\hMpc$. For $\Sigsm=15\hMpc$ and $\Sigsm=10\hMpc$, the the signal to noise at $k\sim 0.2\ihMpc$ is 24\% and 17\% greater across the line of sight, respectively. Therefore, the remaining discrepancy between the Fisher forecasts and our mock data analysis could be partly due to the missing hexadecapole information in the data analysis \citep[e.g.,][]{Taruya11}.  Including the hexadecapole or designing alternative moments, optimized for such a distribution of signal to noise \scr{may} help to alleviate the tension between Fisher forecasts and current measurements. We leave such investigations to future work.

In summary, we observe significant gains in terms of the agreement between the model and the data,  biases on the measurements, and also a modest improvement on the measurement precisions when using the modified model for the isotropically reconstructed data. 

\section{Conclusion}\label{sec:con}
We have investigated the effects of the details that went in the densify field reconstruction technique by means of propagators of the pre and post reconstructed density field. Using a mock galaxy sample that mimics the clustering and noise level of BOSS CMASS galaxies, we studied the effect of the RSD treatment as well as the effect of the smoothing scales used during the BAO reconstruction.

We found that while the original BAO reconstruction convention that attempts to preserve the RSD signal in the reconstructed field induces a boosted signal on large scales along the line of sight, the BAO reconstruction convention that attempts to remove the RSD signal appears to recover a shallower BAO damping. We estimated the noise level for the two conventions before and after reconstruction and found that the noise levels are almost identical on small scales. On large scales where the anisotropic convention produces a boosted signal along the line of sight, the noise level is boosted as well for this convention; the dominant contribution to noise on large scales would be shot noise and any shot noise relative to the signal in the density field we use to derive the displacement field will be frozen despite the overall factor we multiply. As a result, the two conventions returned almost identical BAO signal-to-noise ratios.

For the noise level of BOSS CMASS, we found that the reconstruction efficiency improves as the smoothing scale decreases from $20\hMpc$; the efficiency saturates near the smoothing scale of $7-10\hMpc$ and decreases below $7\hMpc$. 
We clearly showed that the isotropic convention generates a BAO signal along the line of sight that cannot be modeled by the previously used, simple Gaussian damping model and we present a modified model based on the first order LPT that should closely correspond to the model for the correlation function presented in \citet{White15}. The original Gaussian damping model appeared a worse description for a larger smoothing scale. Both Gaussian damping models appeared to fail as the smoothing scale decreases below $10\hMpc$, probably due to more nonlinear and shot-noise dominated information included when reconstructing the displacement field.

Decreasing the smoothing scale not only improved the BAO signal but also decreased the noise level.  We found that such improvement saturates below the smoothing scale of $7\hMpc$.  In terms of BAO signal to noise, the gain is maximized at the smoothing scale of $7$-$10\hMpc$ for BOSS CMASS; considering the validity of the modified model as a fitting formula for $\Sigsm >\sim 10\hMpc$, we believe that a smoothing scale of $10\hMpc$ would be the optimal for the current data set. 

We incorporated the derived propagator and noise results into the Fisher matrix analysis and predicted constraints on an angular diameter distance and Hubble parameter. When the smoothing scale decreased from $20\hMpc$ to $10\hMpc$,  the predicted precision on $\DA$ and $H$ improved by 16\% and 20\%, respectively. When decreasing the smoothing scale from $15\hMpc$ to $10\hMpc$, the precision on $\DA$ and $H$ improved by 10\% and 12\%, respectively. The derived Fisher estimates were approximately 30\% better than the reported DR11 constraints \citep{DR11}. 

Using the mock data, we tested the effect of using the modified BAO model in the data analysis. The modified model substantially improved the goodness of the fit, decreased biases on the recovered cosmology, and mildly improved the constraints. The improved constraints agree with the Fisher forecasts within  15-20\%. We suspect that the remaining discrepancy between the Fisher forecasts and our measurements could be partly due to the missing hexadecapole information in our analysis; we leave this for future investigation. 

\section*{Acknowledgements}
We are very thankful to Martin White for kindly providing runA simulations. We are also thankful to Patrick McDonald, Angela Burden, and Mariana Vargas Magana for very useful comments.
Hee-Jong Seo is thankful to Berkeley Center for Cosmological Physics for a travel support. 
Hee-Jong Seo's work is supported by the U.S. Department of Energy, Office of Science, Office of High Energy Physics under Award Number DE-SC0014329.
Shun Saito is supported by World Premier International Research Center Initiative (WPI Initiative), MEXT, Japan.

\appendix
\begin{widetext}

\section{BAO reconstruction of biased tracers in redshift space}\label{sec:appa}

Assuming locality of bias in Lagrangian space, \citet{Mat08b} derives an expression of the power spectrum 
of biased objects in redshift space. The validity of the local Lagrangian bias can be found in 
e.g., \citet{Saito14}. To summarize, the Eulerian density field $\dx$ and the power spectrum of biased 
tracers $\Pbias(\bk)$ in redshift space can be written as:
\bea
\dx &=&\int d^3 q\,F[\delta_R(\bq)]\,\delta^3_D[\bx-\bq-\bPsin{\rm s}(\bq)]-1 \label{eq:dbias}\\ 
\Pbias(\bk) &=& \int d^3 q\,\enikq\left [ \indlo\indlt \tF(\lambda_1)
  \tF(\lambda_2) \right.\nnm,\\
&& \left. \llan \eia{[\lambda_1\delta_R(\bq_1)+\lambda_2\delta_R(\bq_2)]
-\bk\cdot[\bPsi^{\rm s}(\bq_1)-\bPsin{\rm s}(\bq_2)]}\rran  -1 \right],\nnm\\
\eea
where $\bq = \bq_1-\bq_2$ is the Lagrangian coordinate, and $\bPsin{\rm s}$ is a displacement field 
in redshift space given by $\bPsin{\rm s}=\bPsin{}+(\hat{z}\cdot \dot{\bPsin{}})\hat{z}/H$. 
$\tF$ is the Fourier transform of the Lagrangian bias functional that depends on the smoothed linear 
density field $\delta_R(\bq)$. From these definitions, \citet{Mat08b} derives
\bea
\Pbias(\bk) &=& \exp \left[ -\{k^2 (1-\mu^2) + k^2 \mu^2 (1+f)^2\}A_{v}+ \mathcal{O}(\PL^{2}) \right] \nnm\\
&& \times \left[ (1+ \llan F' \rran + f\mu^2)^2 \PL(k) + \mathcal{O}(\PL^{2}) \right], 
\label{eq:Pbobs}
\eea
where $A_{v}$ is the linear velocity dispersion, $A_{v} \equiv \int\,dp\,\PL(q)/(6\pi^2)$, and the linear bias 
corresponds to $b \equiv 1 + \llan F' \rran$. 

The full expression of the $\mathcal{O}(\PL^{2})$ at one-loop level can be found in \citet{Mat08b}. 
The observed biased field will additionally include the Poisson shot noise contribution $\pshot(\bx)$, 
i.e., 
\bea
\dx &=&\int d^3q\,F[\delta_R(\bq)]\delta^3_D[\bx-\bq-\bPsin{\rm s}(\bq)]-1 \nnm\\
&&+ \int d^3q\,\pshot(\bq) \delta^3_D[\bx-\bq],\nnm\\
&& 
\eea
so that 
\bea
\llan \pshot(\bq) \pshot(\bq') \rran &=& \frac{1}{\nbar}\delta_D^3(\bq-\bq'),
\eea
where $\nbar$ is the number density of tracers. In Fourier space, we have 
\bea
\dko&=& \int d^3q  \,\enikq \left[\int \frac{d\lambda_{1}}{2\pi} \tF(\lambda_1)\eia{\lambda_1\delta_R(\bq)}
\enia{\bk\cdot\bPsin{\rm s}} -1 \right] \nnm \\
&&+\,\,p_{\rm shot}(\bk),\\
\Pobs(\bk)&=& \Pbias(\bk)+\frac{1}{\nbar}\label{eq:ppobs}.
\eea 
Note that the shot noise contribution is nonzero for $\bk \neq \bk'$ if $\nbar$ has a spacial dependence. 
For simplicity, we ignore  the shot noise contribution. 

Now we aim to derive a similar expression for the reconstructed fields. 
Here we closely follow the derivations in \citet{PadLPT} and \citet{Noh2009} that were taken for 
reconstructing real-space biased density field. We derive the reconstructed power spectrum of 
the biased tracers in redshift space, particularly including the two reconstruction conventions. 
As explained in Section 2.2, the reconstruction generally requires us to first estimate 
the displacement field for the observed galaxies, $\bs$, along the line of sight, after applying 
a smoothing kernel $S(\bk)$ with the correction factors, $\lamo$ and $\kappa$ 
which depend on the convention:
\bea
{\bf s}_{i}(\bk) = (\del_{ij} + \lamo\hat{z}_{i}\hat{z}_{j} ) \left(-i\frac{k_j}{k^2}S(\bk) \frac{\dko}{\kappa} \right). 
\eea
Likewise for the reference particles, we displace them with the correction factors, $\lamt$ and $\kappa$ 
which again depend on the convention:
\bea
{\bf s}_{{\rm s},i}(\bk) = (\del_{ij} + \lamt\hat{z_i}\hat{z_j} ) \left(-i\frac{k_j}{k^2}S(\bk) \frac{\dko}{\kappa} \right). 
\eea
Then the displaced galaxy density and random fields are simply given by 
\bea
\dkd &=& \int d^3 x\,\enikx\,\nnm\\
&&\times \left\{\int d^3 q \,F[\delta_R(\bq)]\delta^3_D[\bx-\bq-\bPsin{\rm s}(\bq)-\bs ] -1  \right\} \nnm\\
&=&\int d^3q \,\enikq \left\{\int  \frac{d\lambda_{1}}{2\pi} \tF(\lambda_1)\eia{\lambda_1\delta_R(\bq)} \enia{\bk\cdot(\bPsin{\rm s} + \bs)} - 1\right\}, \nnm\\
&&\\
\dks &=& \int d^3q \,\enikq ( \enia{\bk\cdot \bs_s} - 1),
\eea
and therefore the reconstructed power spectrum $\Prec$ is expressed as
\bea
\Prec(\bk) &=& \llan (\dkd -\dks)(\dkd -\dks)^*\rran  \nnm \\
&=&\Pdd(\bk)+2\Psd(\bk)+\Pss(\bk) \nnm\\
&=&\int d^3 q\, \enikq  \left [  \llan\int \frac{d\lambda_{1}}{2\pi}\int \frac{d\lambda_{2}}{2\pi} \tF(\lambda_1)
  \tF(\lambda_2) \right. \right.\nnm\\
&& \left. \eia{\lambda_1\delta_R(\bq_1)+\lambda_2\delta_R(\bq_2)-\bk\cdot (\bPsi^s(\bq_1)-\bPsi^{\rm s}(\bq_2) + \bs(\bq_1)-\bs(\bq_2))}  \rran \nnm\\
&&-2 \llan  \int \frac{d\lambda_{2}}{2\pi} \tF(\lambda_2)  \eia{ \lambda_2\delta_R(\bq_2) -\bk\cdot\{\bs_s(\bq_1)-\bs(\bq_2)-\bPsi^{\rm s}(\bq_2)\}  } \rran \nnm\\
&&  + \left. \llan \enia{\bk\cdot(\bs_s(\bq_1)-\bs_s(\bq_2))} \rran 
\right]. \label{eq:Prec}
\eea

\subsection{The Rec-Iso convention}\label{subsec:appa}
In the case of the reconstruction convention of `Rec-Iso' from equation (\ref{eq:recisoq}), 
we correct for the redshift-space distortions as well as for the galaxy bias 
in deriving the displacement fields by dividing the observed density field by $b_e(1+\beta_e \mu^2)$, 
where $b_e$ and $\beta_e$ are our estimates of the true quantities.
Simply assuming $b_{e}=b$ and $\beta_{e}=\beta=f/b$, the correction factors in the `Rec-Iso' 
convention are
\bea
\lamo^{\rm Rec-Iso} & = & f,  \nnm \\
\lamt^{\rm Rec-Iso} & = & 0,  \nnm \\
\kappa^{\rm Rec-Iso} & = & b(1+\beta \mu^{2}). 
\eea
Let us derive individual terms of $\Prec$. 
With a help of the cumulant expansion theorem, the first term in equation (\ref{eq:Prec}) becomes

\bea
\Pdd(\bk) &=&   \int d^3 q\, \enikq \llan \left[ \frac{d\lambda_{1}}{2\pi}\frac{d\lambda_{2}}{2\pi} \tF(\lambda_1)
  \tF(\lambda_2)  \eia{(\lambda_1\delta_R(\bq_1)+\lambda_2\delta_R(\bq_2)
  -\bk\cdot (\bPsi^{\rm s}(\bq_1)-\bPsi^{\rm s}(\bq_2) + \bs(\bq_1)-\bs(\bq_2))}  \right] \rran \nnm\\
&=& \int d^3q\, \enikq \frac{d\lambda_{1}}{2\pi}\frac{d\lambda_{2}}{2\pi} \tF(\lambda_1)  \tF(\lambda_2) \exp\left[\sum_{N=1}^{\infty} \frac{(-i)^N}{N!} \right. \llan \left.\left\{- \lambda_1\delta_R(\bq_1)-\lambda_2\delta_R(\bq_2) +\bk\cdot (\bPsi^{\rm s}(\bq_1) -\bPsi^{\rm s}(\bq_2) + \bs(\bq_1)-\bs(\bq_2))\right\}^N\rran_c \right ] \nnm \\
&=&  \int d^3 q\, \enikq \frac{d\lambda_{1}}{2\pi}\frac{d\lambda_{2}}{2\pi} \tF(\lambda_1)\tF(\lambda_2)
\exp \left[\frac{i^{n_1+n_2+m_1+m_2+l_1+l_2}}{n_1!n_2!m_1!m_2!l_1!l_2!}B^{n_1 n_2}_{m_1 m_2,l_1 l_2}(\bk,\bq) \right], 
\eea
where $B^{n_1 n_2}_{m_1 m_2,l_1 l_2}(\bk,\bq)$ is defined as
\bea
 B^{n_1 n_2}_{m_1 m_2,l_1 l_2}(\bk,\bq) &\equiv& (-1)^{m_1 + l_1}\llan (\lambda_1 \delta_{R,1})^{n_1}  
 (\lambda_2 \delta_{R,2})^{n_2} (\bk\cdot\bPsi^{\rm s}_{1})^{m_1} (\bk\cdot\bPsi^{\rm s}_{2})^{m_2} (\bk\cdot\bso)^{l_1} (\bk\cdot\bst)^{l_2} \rran_{c}. 
\eea
Here we omit $\delta_{R}(\bq_{1})$ as $\delta_{R,1}$ etc. Notice that this expression is exact 
and does not involve any approximations. As a reminder,  $n_{1,2}$ are entries 
contributed from galaxy bias, $m_{1,2}$ are from the nonlinear density field, and $l_{1,2}$ are 
contributed from the process of reconstruction. 
In the following, we approximate the displacement the field, $\bPsi$ up to linear order, 
and calculate all the terms up to tree order. 
At tree-level order, we include terms  $n_1+n_2+m_1+m_2+l_1+l_2 \leq 2$.
Assuming that the smoothing kernel is isotropic, $S({\bf k})=S(k)$ and the smoothed observed 
field is approximated by the linear density field, i.e., 
$\langle \tilde{\delta}_{\rm L}({\bf k})\tilde{\delta}_{\rm obs}({\bf k})\rangle\approx b(1+\beta \mu^{2})P_{\rm L}(k)$, 
it is straightforward to obtain
\bea
\Pdd({\bf k})
&=&\exp \left[ -k^2 (1-\mu^2)\int \frac{dp}{6\pi^2}\PL(p)\{1-S(p)\}^2-k^2 \mu^2 \int \frac{dp}{6\pi^2}\PL(p)\{(1+f)-(1+\lamo)S(p)\}^2\right] \nnm\\
&&\times \left[b(1+\beta\mu^2)-(1+\lamo\mu^2)S(k)\right]^2 P_{\rm L}(k), \nnm\\
\Pss({\bf k}) &=& \exp \left[ -\{k^2 (1-\mu^2)+k^2\mu^2(1+\lamt)^2\}\int \frac{dp}{6\pi^2}\PL(p) S(p)^{2}\right]  (1+\lamt\mu^2)^2 S(k)^{2}\PL(k), \nnm\\
\Psd({\bf k}) &=& -\sqrt{\Pdd({\bf k})\Pss({\bf k})}, 
\eea
where $\kappa = b(1+\beta \mu^{2})$ is used. We thus finally reach 
\bea
\Prec(k,\mu) 
&=& \lbrace \left[ b(1+\beta\mu^2)-(1+\lamo\mu^2)S(k) \right]e^{-k^{2}(1-\mu^2)\Sigddxy^2/4} e^{-k^2\mu^2\Sigddz^2/4}\nnm\\
&+&(1+\lamt\mu^2) S(k) e^{-k^2(1-\mu^2)\Sigssxy^2/4}e^{-k^2\mu^2\Sigssz^2/4}  \rbrace^2\PL(k)\nnm,\\
\label{eq:preciso}
\eea
where we define 
\bea
\Sigddxy^2 &\equiv& 2\int \frac{dp}{6\pi^2}\PL(p)(1-S(p))^2,  \\
\Sigddz^2 &\equiv& 2\int \frac{dp}{6\pi^2}\PL(p)\{(1+f)-(1+\lamo)S(p)\}^2,\\
\Sigssxy^2 &\equiv& 2\int \frac{dp}{6\pi^2}\PL(p) S^2(p),\\
\Sigssz^2 &\equiv& 2(1+\lamt)^2 \int \frac{dp}{6\pi^2}\PL(p) S^2(p). 
\eea
Making a specific choice of `Rec-Iso', i.e., $\lamo=f$ and $\lamt=0$, to follow the convention in \citet{Pad12}, gives $\Sigddxy=\Sigddz/(1+f)=\Sigdd$ and $\Sigssxy=\Sigssz=\Sigss$ yield to
\bea
\Prec(k,\mu) 
&=&b^2 \left[ \{1+ \beta\mu^2(1-S(k))\}e^{-k(1-\mu^2)\Sigdd^2/4} e^{-k^2\mu^2(1+f)^2\Sigdd^2/4}\right.\nnm\\
&+&\left.\frac{1}{b}S(k)\left\{e^{-k^2\Sigss^2/4}- e^{-k(1-\mu^2)\Sigdd^2/4} e^{-k^2\mu^2(1+f)^2\Sigdd^2/4}\right\}\right]^2\PL(k).\nnm\\
\label{eq:precmodel}
\eea
Thus the resulting propagator is 
\bea
C(k, \mu ) &=& \left\{ 1+ \beta\mu^2(1-S(k)) \right\}e^{-k(1-\mu^2)\Sigdd^2/4} e^{-k^2\mu^2(1+f)^2\Sigdd^2/4}\nnm\\
&+&\frac{1}{b}S(k)\left\{e^{-k^2\Sigss^2/4}- e^{-k(1-\mu^2)\Sigdd^2/4} e^{-k^2\mu^2(1+f)^2\Sigdd^2/4}\right\}
\eea
Our investigations suggest that $\Sigdd = \Sigss =\Sigma$ is a reasonable approximation for a certain choice of 
the smoothing lengths, yielding to the simple fitting formula we suggest in this paper, Equation (\ref{eq:CkGmod}): 
\bea
\Prec(k,\mu) 
&=& b^2 \left[ \left\{ 1+ \beta\mu^2(1-S(k)) \right\}e^{-k(1-\mu^2)\Sigma^2/4} e^{-k^2\mu^2(1+f)^2\Sigma^2/4}+\frac{1}{b}S(k)e^{-k^2\Sigma^2/4}\left\{1-e^{-k^2\mu^2(2f+f^2)\Sigma^2/4}\right\} \right] ^2\PL(k). \nnm\\
\eea
The second term will contribute for large $k$ and $\mu$, while the first term dominates for small $k$ or $\mu$.
Going back to Eq. \ref{eq:precmodel}, even without assuming $\Sigdd = \Sigss =\Sigma$, we can derive the modified Gaussian damping model Eq \ref{eq:CkGmod} when the second term is small:
\bea
\Prec(k,\mu) \approx b^2 \lbrace  \left[ 1+ \beta\mu^2(1-S(k)) \right]e^{-k(1-\mu^2)\Sigma^2/4} e^{-k^2\mu^2(1+f)^2\Sigma^2/4}\rbrace^2\PL,
\eea
and
\bea
C(k,\mu) \approx  \left[ 1+ \beta\mu^2(1-S(k)) \right]e^{-k(1-\mu^2)\Sigma^2/4} e^{-k^2\mu^2(1+f)^2\Sigma^2/4}\label{eq:CkGmodLPT}
\eea
For the smoothing scales we consider in this paper, indeed the second term is negligible.

\subsection{Rec-Ani limit}\label{subsec:appt}
In the 'Rec-Iso' case,
\bea
\lamo^{\rm Rec-Ani} & = & \lamt^{\rm Rec-Ani} = \frac{f-\beta}{1+\beta},  \nnm \\
\kappa^{\rm Rec-Ani} & = & b. 
\eea
While the derivation is omitted, with $\kappa^{\rm Rec-Ani} =  b$, it gives
\bea
\Pdd
&=& \exp \lbrace -k^2(1-\mu^2) \int \frac{dp}{6\pi^2} \PL(p) \left[ \left( 1-S(p) (1+\frac{1}{5}\beta)\right)^2 \right. \left.+ S(p)^2 \frac{8}{175}\beta^2\right] -k^2\mu^2 \int \frac{dp}{6\pi^2} \PL(p)  \left[\left( (1+f)-(1+\lamo)S(p) (1+\frac{3}{5}\beta) \right)^2 \right. \nnm \\
&&+\left.S(p)^2\frac{12}{175}\beta^2 \right]  \rbrace\nnm \\
&\times& (1+\beta\mu^2)^2\PL(k)\left[b-(1+\lamo\mu^2)S(k)\right]^2, \nnm\\
\eea
and
\bea
\Pss
&=&\exp \lbrace -k^2(1-\mu^2) \int \frac{dp}{6\pi^2} \PL(p) \left[ \left(S(p) (1+\frac{1}{5}\beta)\right)^2 \right. \left.+ S(p)^2 \frac{8}{175}\beta^2\right]-k^2\mu^2 \int \frac{dp}{6\pi^2} \PL(p)  \left[\left( (1+\lamt)S(p) (1+\frac{3}{5}\beta) \right)^2 \right. \nnm \\
&&+\left. S(p)^2\frac{12}{175}\beta^2 \right] \rbrace\nnm \\
&\times& (1+\beta\mu^2)^2\PL(k)\left[(1+\lamt\mu^2)S(k) \right]^2,\nnm\\
&& \\
\eea

With $\lamo^{\rm Rec-Ani}$, we can define
\bea
\Sigddxy^2 &\equiv& 2\int \frac{dp}{6\pi^2}\PL(p)\{(1-S(p) (1+\frac{1}{5}\beta))^2+S(p)^2\frac{8}{175}\beta^2\}\\
\Sigddz^2 &\equiv& 2 \int \frac{dp}{6\pi^2}\PL(p)\{\{(1+f)-(1+f)(1+\frac{3}{5}\beta)/(1+\beta)S(p)\}^2+S(p)^2\frac{12}{175}\beta^2\} \\
\Sigssxy^2 &\equiv& 2\int \frac{dp}{6\pi^2}\PL(p)  S^2(p)\{(1+\frac{1}{5}\beta)^2+\frac{8}{175}\beta^2\}\\
\Sigssz^2 &\equiv& 2 \int \frac{dp}{6\pi^2}\PL(p) S^2(p) \{\{(1+f)(1+\frac{3}{5}\beta)/(1+\beta)\}^2 +\frac{12}{175}\beta^2\}.
\eea

The power spectrum for the reconstructed field is obtained as
\bea
\Prec(k,\mu)&=&b^2(1+\beta\mu^2)^2 \left[ \left\{ 1-\frac{1}{b}\left(1+\frac{f-\beta}{1+\beta} \mu^2\right)S(k) \right\}
e^{-k(1-\mu^2)\Sigddxy^2/4} e^{-k^2\mu^2\Sigddz^2/4}\right.\nnm\\
&&+\left.\frac{1}{b}\left(1+\frac{f-\beta}{1+\beta}\mu^2\right) S(k)e^{-k^2(1-\mu^2)\Sigssxy^2/4} e^{-k^2\mu^2\Sigssz^2/4}\right]^2\PL(k).
\label{eq:precanl}
\eea
Rearranging some terms and approximating $\Sigdd=\Sigddxy=\Sigddz/(1+f)$ and $\Sigss=\Sigssxy=\Sigss/(1+f)$ gives
\bea
\Prec(k,\mu)&=& b^2(1+\beta\mu^2)^2 \left[ e^{-k(1-\mu^2)\Sigdd^2/4} e^{-k^2\mu^2(1+f)^2\Sigdd^2/4}\right.\nnm\\
&&+\left.\frac{1}{b}\left(1+\frac{f-\beta}{1+\beta}\mu^2\right) S(k)       \{  e^{-k^2(1-\mu^2)\Sigss^2/4} e^{-k^2\mu^2(1+f)^2\Sigss^2/4} -   e^{-k^2(1-\mu^2)\Sigdd^2/4} e^{-k^2\mu^2(1+f)^2\Sigdd^2/4} \}    \right]^2\PL(k). \nnm\\
\eea
Again, for the smoothing scales we consider in this paper, the second term is negligible, giving
\bea
\Prec(k,\mu)&\approx& b^2(1+\beta\mu^2)^2 \left[ e^{-k(1-\mu^2)\Sigma^2/4} e^{-k^2\mu^2(1+f)^2\Sigma^2/4}  \right]^2\PL(k). 
\label{eq:precreal}
\eea
and
and 
\bea
C(k,\mu) \approx  \left[ 1+ \beta\mu^2 \right]e^{-k(1-\mu^2)\Sigma^2/4} e^{-k^2\mu^2(1+f)^2\Sigma^2/4},\label{eq:CkGLPT}
\eea
which returns the form in Eq. \ref{eq:CkG}.

Note that the real-space limit ($f=0$) in both conventions Eq.~\ref{eq:preciso} and Eq.~\ref {eq:precanl} reduce to the form derived in \citet{Noh2009}. 
\section{Modeling propagators using the second order LPT}\label{sec:appc}
Following \citet{Mat08b} and \citet{Noh2009}, the pre-reconstructed density field of biased tracers 
in redshift space have the following propagator when expanded to include the next-to-leading order 
(i.e., one-loop) terms:
\bea
C(k,\mu) 
&=&\frac{1}{b\PL}\left[(1+f\mu^2)\PL+\frac{5}{21}R_1+\frac{3}{7}R_2+ f\mu^2(\frac{6}{7}R_2+\frac{5}{7}R_1)+\right. \nnm\\
&&\left. (b-1)\left(\PL+\frac{3}{7}(R_1+R_2)+\frac{6}{7}f\mu^2(R_1+R_2) \right)\right] 
\exp  \left[ -\left\{k^2 (1-\mu^2)+k^2\mu^2(1+f)^2\right\} A_{v}\right],
\eea
where 
\bea
R_1(k)&=&\frac{k^3}{(2\pi)^2}\PL(k)\int_0^\infty dr\,\PL(kr) \int_{-1}^{1}d\mu \frac{r^2(1-\mu^2)^2}{1+r^2-2r\mu}\label{eq:R-1}\\
R_2(k)&=&\frac{k^3}{(2\pi)^2}\PL(k)\int_0^\infty dr\,\PL(kr) \int_{-1}^{1}d\mu \frac{(1-\mu^2)r\mu(1-r\mu)}{1+r^2-2r\mu}\label{eq:R-2}
\eea
\citet{Noh2009} shows that the reconstructed density field of biased tracers in real space have 
the following propagator:
\bea
C(k) &=& \frac{\llan \tilde{\delta}_L\tilde{\delta}_{\rm rec} \rran}{b\PL(k)}\nnm\\
&=&\frac{1}{b\PL}\left[ \left\{\PL(k)(1-S(k))+\frac{5}{21}R_1+\frac{3}{7}R^{(d)}_2 + (b-1)\left(\PL(k)+\frac{3}{7}(R_1+R_2) \right) \right\} e^{-k^2\Sigdd^2/4}+
\PL S e^{-k^2\Sigss^2/4} \right],
\eea
where the superscript `(d)' in $R^{(d)}$ means using $\PL (1-S)$ inside the integral in 
Equations (\ref{eq:R-1}) and (\ref{eq:R-2}). 
Removing the higher-order contributions involving $R$ returns the $f=\beta=0$ case in 
Equation (\ref{eq:precanl}).
In this paper, we expand their result to the redshift space and derive the 2nd-order LPT 
propagator model for the biased tracers in redshift space. 
Here we only consider the `Rec-Iso' case just for simplicity: 
\bea
C(k,\mu) &=&\frac{1}{b\PL}\left[\left[\left\{(1+f\mu^2)-(1+\lamo\mu^2)S(k)\right\}\PL(k)
+\frac{5}{21}(1+3f\mu^2)R_1+\frac{3}{7}R^{(d)}_2+ f\mu^2\left(\frac{6}{7}R^{(d)}_2+\frac{3}{7}\lamo R^{(s)}_1-\frac{3}{7}\lamo R^{(s)}_s\right)+\right. \right. \nnm\\
&&-\frac{3}{7}\mu^4f\lamo(R^{(s)}_1+2R^{(s)}_2)\\
&&\left. \left. (b-1)\left\{\PL+\frac{3}{7}(R_1+R_2)+\frac{6}{7}f\mu^2(R_1+R_2) \right\}\right]e^{-k^2\Sigdd^2/4}+
(1+\lamt\mu^2)\PL(k)S(k) e^{-k^2\Sigss^2/4} \right],
\eea
where the superscript `(s)' in $R^{(s)}$ means using $\PL(k)S(k)$ inside the integral in 
Equations (\ref{eq:R-1}) and (Eq. \ref{eq:R-2}).With $\lamo=f$ and $\lamt=0$, this gives
\bea
C(k,\mu) &=&\frac{1}{b\PL}\left[\left[(1+f\mu^2)\PL(1-S)+\frac{5}{21}(1+3f\mu^2)R_1+\frac{3}{7}R^{(d)}_2+ 
f\mu^2(\frac{6}{7}R^{(d)}_2+\frac{3}{7}f R^{(s)}_1-\frac{3}{7}f R^{(s)}_s)+\right.\right. \nnm\\
&&-\frac{3}{7}\mu^4f^2(R^{(s)}_1+2R^{(s)}_2)\\
&&\left. \left. (b-1)\left\{\PL+\frac{3}{7}(R_1+R_2)+\frac{6}{7}f\mu^2(R_1+R_2) \right\}\right]e^{-k^2\Sigdd^2/4}+
\PL S e^{-k^2\Sigss^2/4} \right],
\eea
\end{widetext}

\begin{figure*}[h]
 \centering
\includegraphics[width=0.8\linewidth]{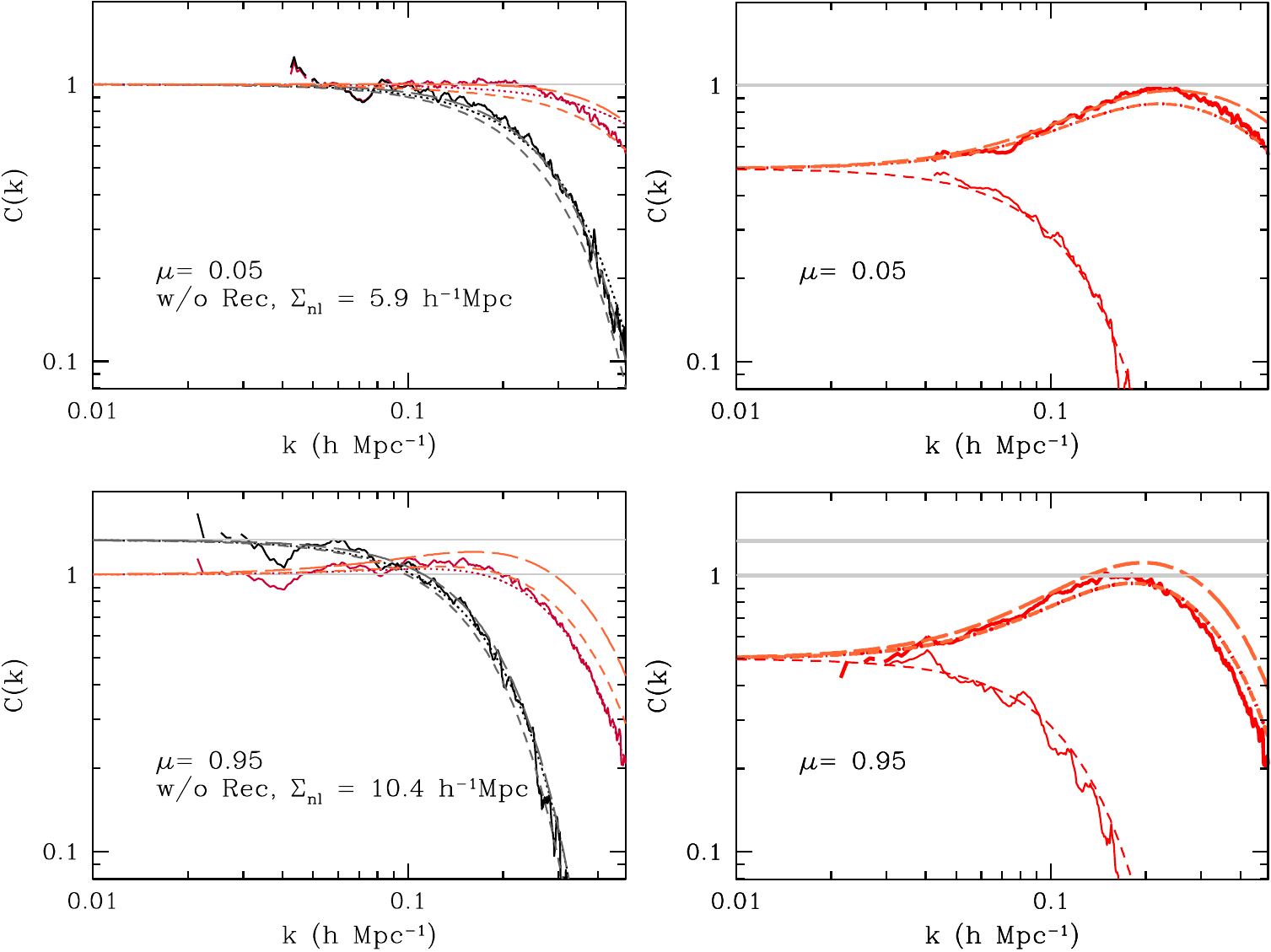}
\caption{Comparison between models and the measured propagators. Solid lines: measured propagators with smoothing length of $14 \hMpc$. Dotted lines: our Gaussian model based on $C(k)$ at $k=0.3\ihMpc$. Short dashed lines: the first order LPT model . Long dashed lines: including the second order terms.  The right-hand panels again show the corresponding individual density fields, i.e., the displaced random field  ($-\delta_s$, the \scr{thinner} curves peaking at low $k$) and 
displaced galaxy density field ($\delta_d$, the \scr{thicker} curves peaking at high $k$) that forms the reconstructed density 
field (i.e., in the left-hand panels) after mutual addition. Note that when we include the second order terms, we can describe $C(k)$ before reconstruction (black lines) better than when using the first order term alone, while the first order model alone is a better fit to $C(k)$ after reconstruction. We find that our simple models based on Eq. \ref{eq:CkG} and \ref{eq:CkGmod} (dotted lines) describe the measured $C(k)$ fairly accurately before and after reconstruction. }\label{fig:figLPTa}
\end{figure*}

Figure~\ref{fig:figLPTa} compares the measured propagators after isotropic reconstructions with $\Sigsm = 14\hMpc$ (i.e., the same measurement as in Figure \ref{fig:figAni}) with the first and the second order LPT models.
Here the dotted lines are our Gaussian model based on $C(k)$ at $k=0.3\ihMpc$, i.e, the same as the dotted lines in Figure \ref{fig:figAni}, while the short dashed lines are from the first order LPT model and the long dashed lines are including the second order LPT terms. We find that including the second order terms improves the agreement between the measured $C(k)$ and the model before reconstruction (black lines), while it worsens the agreement after reconstruction along the line of sight. We suspect that the assumptions such as the smoothed nonlinear density field being the smoothed linear density field may breaks down when we consider higher order terms. Note that our simple models based on Eq. \ref{eq:CkG} and \ref{eq:CkGmod} (dotted lines) describe the measured $C(k)$ fairly accurately before and after reconstruction. We find that the second term in Eq. \ref{eq:precmodel} is indeed negligible and we conclude that our modified Gaussian model is a good description of the post-reconstruction propagators.

For the redshift of $z=0.55$ of the mock we consider in this paper, we expect nonlinear damping of $ \int \frac{dp}{6\pi^2}\PL(p)=6.47\hMpc$ across the line of sight and $(1+f) \int \frac{dp}{6\pi^2}\PL(p)=11.26\hMpc$ along the line of sight. These values are slightly larger than the Gaussian model values we quote based on $C(k)$ at $k=0.3\ihMpc$, i.e., $\Signl=5.9\hMpc$ across the line of sight (to be exact, at $\mu=0.05$) and $10.4\hMpc$ along the line of sight  (to be exact, at $\mu=0.95$), which agrees with the first order approximation slightly overestimating the damping in Figure \ref{fig:figLPTa}.  After reconstruction, we predict $\Sigdd = 2.52\hMpc$, $(1+f)\Sigdd=4.38\hMpc$, $\Sigss=5.4\hMpc$, $\Sigsd=4.21\hMpc$ with $\Sigsm=10\hMpc$.

With $\Sigsm=14\hMpc$, we predict $\Sigdd = 3\hMpc$, $(1+f)\Sigdd=5.22\hMpc$, $\Sigss=5.05\hMpc$, $\Sigsd=4.15\hMpc$. Finally with $\Sigsm=20\hMpc$, we predict $\Sigdd = 3.52\hMpc$, $(1+f)\Sigdd=6.12\hMpc$, $\Sigss=4.63\hMpc$, $\Sigsd=4.11\hMpc$.

Based on Eq. \ref{eq:precmodel}, when the second term Eq. \ref{eq:precmodel} is small, which seems to be the case, we expect that $(1+f) \Sigdd$ should agree with the measured $\Signl$ values along the line of sight after reconstruction. We found 5.1, 5.6, and $6.2\hMpc$ for $\Sigsm = 10$, 14, and $20\hMpc$ from \S~\ref{subsec:signalsm}, showing a good agreement for larger smoothing lengths compared to for smaller smoothing lengths.

\end{document}